\documentclass[11pt, oneside]{article}   	% use "amsart" instead of "article" for AMSLaTeX format
\pdfoutput=1 % if your are submitting a pdflatex (i.e. if you have images in pdf, png or jpg format)
\usepackage{jheppub}

\usepackage{amsmath, amsthm, graphicx,amssymb,subfigure,bbm}
\usepackage[all]{xy}
\preprint{DIAS-STP-15-09 }

\author[]{Veselin G. Filev,}
\author[]{Denjoe O'Connor,}
\affiliation[]{  School of Theoretical Physics,\\ 
       Dublin Institute for Advanced Studies, \\
       10 Burlington Road, 
       Dublin 4, Ireland.}
\emailAdd{vfilev@stp.dias.ie}
\emailAdd{denjoe@stp.dias.ie}

\abstract{We study the maximally supersymmetric BFSS model at finite
  temperature and its bosonic relative. For the bosonic model in $p+1$
  dimensions, we find that it effectively reduces to a system of
  gauged Gaussian matrix models. The effective model captures the low
  temperature regime of the model including one of its two phase
  transitions.  The mass becomes $p^{1/3}\lambda^{1/3}$ for large $p$,
  with $\lambda$ the 'tHooft coupling. Simulations of the bosonic-BFSS
  model with $p=9$ give $m=(1.965\pm .007)\lambda^{1/3}$, which is also
  the mass gap of the Hamiltonian.  We argue that there is no `sign'
  problem in the maximally supersymmetric BFSS model and perform
  detailed simulations of several observables finding excellent
  agreement with AdS/CFT predictions when $1/\alpha'$ corrections are
  included.}

\title{The BFSS model on the lattice}
\begin{document}
\maketitle
%\section{}
%\subsection{}

\section{Introduction}
The leading candidate for a non-perturbative formulation of M-theory
is expected to be the infinite matrix size limit of a matrix model of
some kind.  One such proposal is the BFSS model
\cite{Banks:1996vh,Townsend:1995kk,Susskind:1997cw} which was
conjectured to capture the entire dynamics of M-theory\footnote{With a
  periodically identified lightlike direction for finite matrix size.}.
Relatives of this model such as the BMN model \cite{Berenstein:2003gb}
or models derived from the ABJM model \cite{Aharony:2008gk} are also
considered possible viable candidates for such a non-perturbative
formulation.  All of these conjectured formulations of M-theory are
regularised versions of the supermembrane and are matrix quantum
mechanical systems. They are based on the matrix regularisation of
membranes introduced by Hoppe \cite{Hoppe:PhDThesis1982} and extended
to the supermembrane in \cite{Townsend:1995kk} and \cite{de
  Wit:1988ig}.

In the second half of the paper we focus on the BFSS model. However,
we dedicate the earlier sections of the paper to a careful study of
the Hoppe regulated bosonic membrane which is the bosonic part of the
BFSS model.  This model is also of separate interest since it is the
high temperature limit of maximally supersymmetric $1+1$ dimensional
$SU(N)$ Yang-Mills (or matrix string theory) compactified on a circle
when the fermions decouple due to their anti-periodic boundary
conditions at finite temperature. The model has been studied already
both theoretically \cite{Mandal:2011hb} (see also ref. \cite{Mandal:2009vz})  and numerically \cite{Azuma:2014cfa, Catterall:2010fx} and our results are
in accord with these studies. The zero temperature supersymmetric gauge theory has been studied in a hamiltonian light cone context in ref.~\cite{Hiller:2005vf}.

We then continue our study with the BFSS model. This model first
emerged as ${\cal N}=16$ supersymmetric Yang-Mills quantum mechanics
\cite{Baake:1984ie,Flume:1984mn,Claudson:1984th}, later it arose as
the 11-dimensional supermembrane in Hoppe's regularization and most
recently as the BFSS model \cite{Banks:1996vh}, a candidate for a
non-perturbative formulation of M-theory, and it also describes the low energy
effective theory of a system of $D0$-branes \cite{Witten:1995im}.
A lattice version of the model which preserves eight of the sixteen supersymmetries was presented in \cite{Kaplan:2005ta}.

Our lattice regularisation of the model follows Catterall and Wiseman
\cite{Catterall:2008yz}. In this formulation it is clear that the
Pfaffian obtained when the Fermions are integrated out can in general
be complex. However, we find that the phase of the Pfaffian is
restricted to a narrow band so that $\cos(\Theta_{pf})\sim1$. There is
therefore no real phase or `sign problem' as far as simulations of the
model are concerned.  We simulate the model using a rational hybrid
monte carlo algorithm (RHMC) and find excellent agreement with results
reported in \cite{Catterall:2008yz,Anagnostopoulos:2007fw, Hanada:2008ez, Kadoh:2015mka} though our values for the energy are
slightly higher than those of
\cite{Anagnostopoulos:2007fw,Kadoh:2015mka} but in excellent agreement
with predictions of AdS/CFT when leading $1/\alpha'$ corrections are
included. Those interested primarily in the supersymmetric model can
skip to section \ref{supersymmetric section} for our discussion of the
model and results, consulting section \ref{The BFSS model} for the
continuum model and our notation.

The principal results of this paper are:
\begin{itemize}
\item{}We perform monte carlo simulations of the bosonic BFSS model
  and measure the two point correlation function, the mass gap and the
  eigenvalue distribution of each matrix. All fit with Gaussian matrix
  quantum mechanics with the same mass as that found from the
  correlation function.
\item{} We derive an effective description of the bosonic model using a $1/p$ 
expansion where $p$ is the number of matrices. The description is in terms of 
$p$ massive scalar fields that are gauged under the adjoint action of $SU(N)$ 
but are otherwise free scalar fields\footnote{The relevant results here
  were previously obtained in \cite{Mandal:2011hb} which we became aware of these
  after writing this article.}
\item{}The effective model reproduces the one of two phase transitions (see
  \cite{Kawahara:2007fn}) of the full model with surprising
  precision\footnote{The model can also be interpreted as the high temperature
  description of the maximally supersymmetric $1+1$ dimensional
  Yang-Mills compactified on a circle. In this setting its two
  transitions are the high temperature limit of the black hole black
  string transition in the dual gravity model.}.
\item{}We study the maximally supersymmetric BFSS model and present
  arguments showing that though the lattice model in general has a
  complex phase, it is only the cosine of this phase that enters in
  simulations and the model is such that the angle is restricted to
  regions where the cosine is positive hence there is no sign problem
  in the full model.
\item{}We simulate the model using a Fourier accelerated rational
  hybrid monte carlo algorithm confirming results found earlier by
  other groups and find excellent agreement with predictions of
  AdS/CFT when subleading $1/\alpha'$ corrections are included.
\end{itemize}

The simulations we report provide a useful test of our code as we
proceed to examine further observables and the inclusion of
longitudinal M5-branes or equivalently $D4$-branes. Our goal being the
Berkooz-Douglas matrix model holographically dual to the backreacted
D0/D4 brane intersection. Such studies can provide a solid test of the
AdS/CFT correspondence with flavour degrees of freedom, which is a
widely used tool for non-perturbative analysis in flavour dynamics.

The layout of the paper is as follows: In section 2 we introduce the
BFSS model and continue in section 3 with a study of the bosonic part
of the model describing our lattice discretisation and hybrid monte
carlo algorithm.  We then present our numerical results for this
bosonic model and continue in section 3.5 to develop an expansion for
the model in terms of the inverse of the number of matrices. This
leads us to introduce our effective Gaussian model and compare it with
the full model. In section 4 we return to the supersymmetric model and
present our lattice discretisation of this model. We give a discussion
of the Pfaffian phase of the model and present our results.  The paper
ends with a discussion of our results in section 5.

\section{The BFSS model}
\label{The BFSS model}
The BFSS matrix model is a one dimensional supersymmetric Yang-Mills
theory naturally arising in type IIA superstring theory as a low
energy effective description of D0-branes. It is conjectured that in
the limit of a large number of D0-branes, $N$, the model is equivalent to
uncompactified eleven dimensional $M$-theory \cite{Banks:1996vh} while
for finite $N$ it corresponds to $M$-theory compactified on a
light-like circle \cite{Susskind:1997cw}. The easiest way to obtain the
BFSS matrix model is via dimensional reduction of ten dimensional
supersymmetric Yang-Mills theory down to one dimension. The resulting
reduced ten dimensional action is given by \cite{Polchinski}:
\begin{equation}\label{BFSS 10Mink}
S_M =\frac{1}{g^2}\int dt \,{\rm Tr}\left\{\frac{1}{2}({\cal D}_0X^i)^2 +\frac{1}{4}[X^i,X^j]^2-\frac{i}{2}\Psi^T C_{10}\,\Gamma^0D_0\Psi +\frac{1}{2}\Psi^T C_{10}\,\Gamma^i[X^i,\Psi]\right\}\ ,
\end{equation}
where $\Psi$ is a thirty two component Majorana--Weyl spinor,
$\Gamma^\mu$ are ten dimensional gamma matrices and $C_{10}$ is the
charge conjugation matrix satisfying $C_{10} \Gamma^{\mu}C_{10}^{-1} =
-{\Gamma^{\mu}}^T$. We take a representation for $\Gamma^\mu$ in terms
of nine dimensional (euclidian) gamma matrices $\gamma^i$:
\begin{eqnarray}\label{gammas}
\Gamma^i &=& \gamma^i\otimes \sigma_1\ , ~~~~{\rm for }~ i = 1,\dots ,9\ , \nonumber \\
\Gamma^0 &=& 1_{16}\otimes i\sigma_2\ , \nonumber \\
C_{10} &=&C_{9}\otimes i\sigma_2\ ,
\end{eqnarray}
where $C_9$ is the charge conjugation matrix in nine dimensions satisfying $C_{9} \gamma^{i}C_{9}^{-1} = {\gamma^{i}}^T$  and $\sigma_i$ are the Pauli matrices. 
With the following choice for the Majorana--Weyl spinor:
\begin{equation}\label{Weyl}
\Psi = \psi \otimes \left(\begin{array}{c} 1 \\ 0 \end{array}\right)\ ,
\end{equation}
where $\psi$ is a sixteen component $Spin(9)$ Majorana fermion and
Wick rotating to Euclidean time, we obtain the Euclidean action:
\begin{equation}\label{BFSS-9D}
S_E =\frac{1}{g^2}\int d\tau \,{\rm Tr}\left\{\frac{1}{2}(D_\tau X^i)^2 -\frac{1}{4}[X^i,X^j]^2+\frac{1}{2}\psi^T C_{9}\,D_\tau\psi -\frac{1}{2}\psi^T C_{9}\,\gamma^i[X^i,\psi]\right\}\ ,
\end{equation}
which, as is manifest, is $Spin(9)$ invariant. 
Note: we have not imposed any restriction on the nine
 dimensional spinor basis. For example if we choose $\gamma^i$ to be
 in the Majorana representation (where the nine $\gamma^i$ are taken
 to be real and symmetric) then $C_9=1_{16}$ and we arrive at the more
 standard form for the action (\ref{BFSS-9D}). However, we are interested
 in a basis in which the discrete theory is free of fermion doubling,
 which can be achieved by taking a basis \cite{Catterall:2008yz} in
 which $C_9=1_{8}\otimes \sigma_1$. We will return to this in section
 \ref{supersymmetric section}, where we consider the discretisation of
 the full BFSS matrix model.

\section{Bosonic BFSS on the lattice}
\label{bosonic section}
In this section we focus on the bosonic part of the action
(\ref{BFSS-9D}) given by:
\begin{equation}\label{BosS}
S_b = \frac{1}{g^2}\,\int_0^{\beta}dt\,{\rm tr}\left\{\frac{1}{2}({\cal D}_t {X^i})^2-\frac{1}{4}[X^i,X^j]^2\right\}\ .
\end{equation}
The zero temperature model was introduced by Hoppe
\cite{Hoppe:PhDThesis1982} as a gauge fixed and regulated description
of membranes. The model has also been extensively studied in the
literature\footnote{The model is also the high temperature limit of
  $1+1$ dimensional ${\cal N}=8$ supersymmetric Yang-Mills on ${\bf
    R}\times S^1$ where $\beta$ is now the period of the $S^1$ and the
  fermions drop out due to their anti-periodic boundary conditions at
  finite temperature.}.  It has been simulated for a first time in refs.~\cite{Aharony:2004ig, Aharony:2005ew}, where certain aspects of the full model were described in terms of simple Gaussian model.  Its phase structure at finite temperature has
been explored in~\cite{Kawahara:2007fn, Mandal:2011hb, Azuma:2014cfa}, where the authors
found that as the temperature is decreased the model first undergoes a
2nd order deconfining-confining phase transition into a phase with
non-uniform but gapless distribution for the holonomy.  As the
temperature is further decreased there is a 3rd order transition to a
gapped holonomy with a quadratic decrease in the internal energy to a
constant value for lower temperatures. The high temperature expansion
of the model was developed in~\cite{Kawahara:2007ib}. In what follows
we will reproduce the main results of~\cite{Kawahara:2007fn, Mandal:2011hb}
and elaborate on the properties of the theory at zero temperature. In
particular we will provide evidence that the low temperature phase of
the model has an effective description in terms of a free massive
scalar which captures many of the finite temperature features of the model
including one of its two phase transitions.

\subsection{Discretisation}
We discretise time to $\Lambda$ sites $t_n = a\, n$, ($n=0,\dots,\Lambda-1$), 
where the lattice spacing is $a=\beta/\Lambda$,
and impose periodic boundary conditions identifying the 
point $t_{\Lambda}=\Lambda a =\beta$ with the point $0$. 
To discretise the covariant derivative ${\cal D}_t$ we
define the transporter fields:
\begin{equation}\label{linkU}
U_{n,n+1} ={\cal P}\exp\left[i\int_{n a}^{(n+1)a}dt\,A(t)\right]\ ,
\end{equation}
where ${\cal P}$ denotes a path ordered product. Let us consider for a moment the pure derivative part of ${\cal D}_t$.  On the lattice we have:
\begin{equation}
 \partial_t X_n^i\rightarrow\frac{X_{n+1}^{i}-X_n^i}{a}\ .
\end{equation}
%:
To make the above expression gauge covariant we have to transport back the field at $t_{n+1}$ to $t_n$. For the discrete version of the covariant derivative, we obtain:
\begin{equation}\label{cov_dev}
{\cal D}_t\rightarrow \frac{U_{n,n+1}X_{n+1}^iU_{n+1,n} -X_n^i}{a}\ ,
\end{equation}
where $U_{n+1,n} = U_{n,n+1}^{\dagger}$. Using equation (\ref{cov_dev}) for the discrete bosonic action we obtain:
\begin{equation}\label{Sb-discr}
S_b = N\sum_{n=0}^{\Lambda-1}{\rm tr}\left\{-\frac{1}{a}\,X_n^iU_{n,n+1}X^i_{n+1}U_{n,n+1}^{\dagger}+\frac{1}{a}\, (X_n^i)^2-\frac{a}{4}\,[X_n^i,X_n^j]^2\right\}\ ,
\end{equation}
where without loss of generality we have taken $g =\frac{1}{\sqrt
  {N}}$.\footnote{This can always be arranged by an appropriate
  rescaling of the matrices and the time coordinate and $\beta$
  becomes the dimensionless temperature parameter
  $\frac{\lambda^{1/3}}{T}$ with $\lambda=g^2N$ the 't Hooft
  coupling.} The action $S_b$ can be written in a much simpler form by
using the ${\rm U}(n)^\Lambda$ gauge symmetry of the model. Indeed, at
each lattice site we have a local ${\rm U}(N)$ symmetry. Using that
symmetry we can perform the transformation:
\begin{eqnarray}
&&{X'}_0^i  =X_0^i\ ,\\ 
 &&{X'}_1^i = U_{0,1}\,X_1^i\,U_{0,1}^{\dagger}\ , \nonumber \\
 &&{X'}_2^i = (U_{0,1}U_{1,2})\,X_2^i\,(U_{0,1}U_{1,2})^{\dagger}\ ,\nonumber\\
 &&\dots\nonumber \\
&&{X'}_{\Lambda-1}^i = (U_{0,1}U_{1,2}\dots U_{\Lambda-2,\Lambda-1})\,X_{\Lambda-1}^i\,(U_{0,1}U_{1,2}\dots U_{\Lambda-2,\Lambda-1})^{\dagger}\nonumber \, 
\end{eqnarray}
introducing the notation ${\cal W} =(U_{0,1}U_{1,2}\dots U_{\Lambda-2,\Lambda-1}U_{\Lambda-1,0})$ for the bosonic action (\ref{Sb-discr}) we obtain:
\begin{equation}\label{Sb-discr-1}
S_b= -\frac{1}{a}N{\rm tr}\left\{\sum_{n = 0}^{\Lambda-2}{X'}_n^i{X'}^i_{n+1}+{X'}_{\Lambda-1}^i{\cal W}\,{X'}_0^i{\cal W}^{\dagger}\right\}+N\sum_{n=0}^{\Lambda-1}{\rm tr}\left\{\frac{1}{a}\, ({X'}_n^i)^2-\frac{a}{4}\,[{X'}_n^i,{X'}_n^j]^2\right\}\ .
\end{equation}
The unitary matrix ${\cal W}$ has the decomposition ${\cal W}=VD V^{\dagger} $, where $D ={\rm diag}\{e^{i\theta_1},\dots,e^{i\theta_N}\}$ is a diagonal unitary matrix and $V$ is a unitary. But the action (\ref{Sb-discr-1}) has the residual symmetry ${X'}_n^i \rightarrow V {X'}_n^i V^{\dagger} $, which we can use to diagonalise ${\cal W}$. Furthermore, it has an additional symmetry ${X'}_n^i \rightarrow h_n {X'}_n^i h_n^{\dagger} $, where $h_n$ is a diagonal unitary matrix, which we can use to ``distribute'' the diagonal matrix $D$ among all of the hop terms. Indeed, defining the matrix $D_\Lambda ={\rm diag}\{e^{i\theta_1/\Lambda},\dots,e^{i\theta_N/\Lambda}\}$, which satisfies $(D_{\Lambda})^{\Lambda}=D$, one can verify that under the transformation: 
\begin{equation}\label{transformation}
{X'}_n^i = (V h_n ) \tilde X_n^i (V h_n)^{\dagger}  ~~~,{\rm where:}~~h_n = (D_{\Lambda})^n\ ,
\end{equation}
the action (\ref{Sb-discr-1}) transforms into:
\begin{equation}\label{Sb-discr-2}
S_b[\tilde X,D_{\Lambda}] = N\sum_{n=0}^{\Lambda-1}{\rm tr}\left\{-\frac{1}{a}\,\tilde X_n^iD_{\Lambda}\tilde X^i_{n+1}D_{\Lambda}^{\dagger}+\frac{1}{a}\, (\tilde X_n^i)^2-\frac{a}{4}\,[\tilde X_n^i,\tilde X_n^j]^2\right\}\ .
\end{equation}
Now let us discuss the measure of the transporter fields $U_{n,n+1}$. The measure can be written as:
\begin{equation}
\prod_{n =0}^{\Lambda-1} {\cal D}U_{n,n+1} = \prod_{n =1}^{\Lambda-1} {\cal D}U_{n,n+1} \, {\cal D}U_{0,1} = \prod_{n =1}^{\Lambda-1} {\cal D}U_{n,n+1} \, {\cal D}{\cal W}\ , 
\end{equation}
where we have used that 
$U_{0,1} = {\cal W}\,(U_{1,2}\dots U_{\Lambda-2,\Lambda-1})^{\dagger}$ and the 
invariance of the measure. But the action (\ref{Sb-discr-2}) 
depends only on the matrix ${\cal W}$ (in fact only on the eigenvalues 
of ${\cal W}$). Therefore the integration over the measure of the 
transporter fields reduces to:
\begin{eqnarray}
&&\int \prod_{n =0}^{\Lambda-1} {\cal D}U_{n,n+1} e^{-S_b[\tilde X,D_{\Lambda}]} =({Vol}\, {\rm U}(N))^{\Lambda-1}\int {\cal D W} e^{-S_b[\tilde X,D_{\Lambda}]}\propto \\
&\propto&\int\prod_{k =1}^Nd\theta_k\,\prod_{l >m}|e^{i\theta_l}-e^{i\theta_m}|^2\,e^{-S_b[\tilde X,D_\Lambda(\theta)]}  \propto \int\prod_{k =1}^Nd\theta_k\,e^{-S_b[\tilde X,D_\Lambda(\theta)]-S_{\rm FP}[\theta]} \nonumber\ ,
\end{eqnarray}
where $S_{\rm FP}[\theta]$ is given by:
\begin{equation}
S_{\rm FP} [\theta]= -\sum_{l\neq m}\ln\left|\sin \frac{\theta_l-\theta_m}{2}\right|\ .
\end{equation}
\subsection{Hybrid Monte Carlo}

To implement the Hybrid Monte Carlo algorithm it is convenient to work in a gauge in which the holonomy matrix is non-trivial at only one link (we choose the one connecting sites zero and $\Lambda-1$). To this end we omit the diagonal matrices $h_n$ in the transformation (\ref{transformation}). The action (\ref{Sb-discr-2}) is then given by:
\begin{equation}\label{Sb-discr-3}
S_b[X,D] = N{\rm tr}\left\{-\frac{1}{a}\,\sum_{n=0}^{\Lambda-2}X_n^iX^i_{n+1}-\frac{1}{a}\,X_{\Lambda-1}^iDX^i_{0}D^{\dagger}+\sum_{n=0}^{\Lambda-1}\left[\frac{1}{a}\, (X_n^i)^2-\frac{a}{4}\,[X_n^i,X_n^j]^2\right]\right\}\ .
\end{equation}
The corresponding Hamiltonian for the molecular dynamics part of the HMC algorithm is then:
\begin{equation}
{\cal H}_{bos} = \frac{1}{2}\sum_{n = 0}^{\Lambda-1}{\rm tr}\, P_n^i.P_n^i + \frac{1}{2}\sum_{l = 0}^{N-1} \,{P_d^l}^2+S_b[X,D(\theta)]+S_{\rm FP}[\theta]\ ,
\end{equation}
where $P_n^i$ are the canonical momenta corresponding to the hermitian matrices $\tilde X_n^i$,  and $p_d^l$ are the canonical momenta associated to the angles $\theta_l$.  Hamilton's equations read:
\begin{eqnarray}
&&\dot P_{n,\,lm}^i = -{\partial{S_b}}/{\partial X_{n,\,ml}^i}\ ,~~~\dot P_d^l = -\partial{S_b}/\partial{\theta_l}\ , \\
 &&\dot X_{n,\,lm}^i = P_{n,\,lm}^i\ ,~~~ \dot \theta_l = P_d^l\ , \nonumber
\end{eqnarray}
where dots denote derivatives are with respect to the Monte Carlo time. Using equation (\ref{Sb-discr-3}), for the corresponding forces we obtain:
\begin{eqnarray}
-{\partial{S_b}}/{\partial X_{0,\,ml}^i} &=&\frac{N}{a}\left(X_{1}^i-2X_0^i+D^{\dagger}X_{\Lambda-1}^iD\right)_{lm}+N a\left[X_0^j,\left[X_0^i,X_0^j\right]\right]_{lm}\ ,\nonumber\\
-{\partial{S_b}}/{\partial X_{n,\,ml}^i} &=&\frac{N}{a}\left(X_{n+1}^i-2X_n^i+X_{n-1}^i\right)_{lm}+N a\left[X_n^j,\left[X_n^i,X_n^j\right]\right]_{lm}~~~{\rm for}~~~n=1,\dots,\Lambda-2\ ,\nonumber \\
-{\partial{S_b}}/{\partial X_{\Lambda-1,\,ml}^i} &=&\frac{N}{a}\left(DX_0^iD^{\dagger}-2X_{\Lambda-1}^i+X_{\Lambda-2}^i\right)_{lm}+N a\left[X_{\Lambda-1}^j,\left[X_{\Lambda-1}^i,X_{\Lambda-1}^j\right]\right]_{lm}\ ,\nonumber \\
-\partial{S_b}/\partial{\theta_l} &=& \frac{2N}{a}\sum_{m = 0}^{N-1}\Re\left(i X_{\Lambda-1\, ml}^iX_{0\, lm}^i e^{i(\theta_l-\theta_m)}\right)+\sum_{m\,,\,m\neq l}\cot\left(\frac{\theta_l-\theta_m}{2}\right)\ ,
\end{eqnarray}
which we implement in the Hybrid Monte Carlo.

\subsection{Phase structure}
\label{phase structure}
\begin{figure}[t] %  figure placement: here, top, bottom, or page
   \centering
   \includegraphics[width=3.9in]{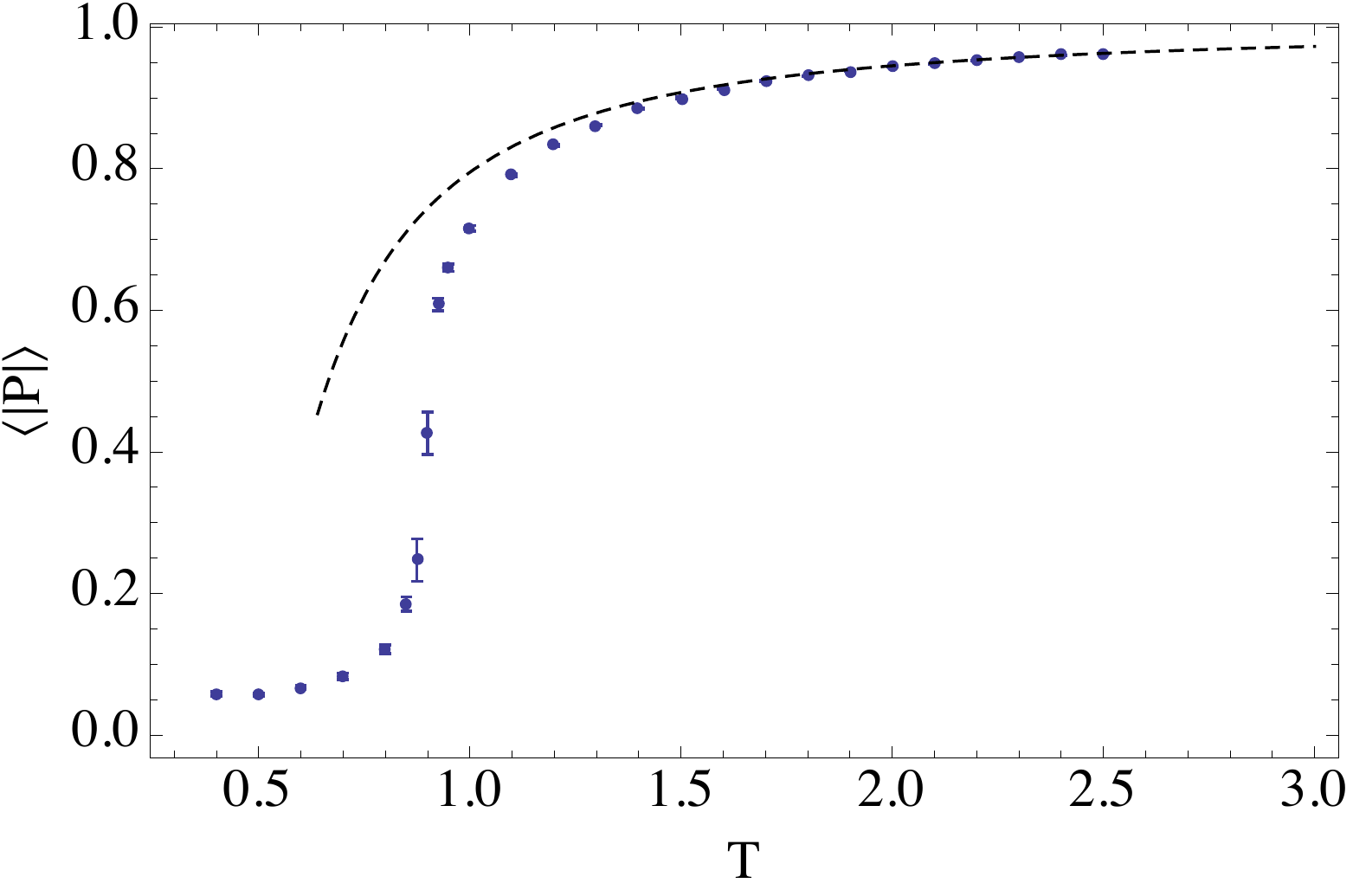} 
 \caption{\small A Plot of the expectation value of the Polyakov loop $|P|$ as a function of the temperature, for $N=16$ and lattice spacing $a\approx 0.05$. One can see that near $T\approx 0.90$ the theory undergoes a second order phase transition. }
   \label{fig:1}
\end{figure}

In this section we reproduce the studies of the phase structure of the
bosonic model originally obtained in~\cite{Kawahara:2007fn}. The
main observables that we focus on are the internal energy of the
system $E$, the ``extent of space'' $\langle R^2\rangle$ and the
Polyakov loop $P$ defined via:
\begin{eqnarray}
E/N^2 &=&\left\langle-\frac{3}{4 N\beta}\int\limits_0^\beta dt\,{\rm Tr}\left([X^i,X^j]^2\right)\right\rangle \ ,\\
\langle R^2\rangle &=&\left\langle\frac{1}{N\beta}\int\limits_0^\beta dt\,{\rm Tr}\left(X^i\right)^2\right\rangle\ . \\
P&\equiv&\frac{1}{N}{\rm Tr} U\ , \\
U&\equiv&{\cal P}\exp\left(i\int\limits_0^\beta d t\,A_0(t)\right)\ ,\label{hol}
\end{eqnarray}
where the holonomy matrix, $U$, is the continuum limit of the link
variable $U_{0,\Lambda}$ defined in equation (\ref{linkU}).  The
expectation value of the Polyakov loop $\langle |P|\rangle$ plays the
r\^ole of an order parameter for the confining-deconfining phase transition 
discussed in~\cite{Kawahara:2007fn}. 
In figure \ref{fig:1} we presented a plot of this order parameter as 
a function of the temperature. The plot is for $N=16$ and lattice 
spacing $a\approx 0.05$. One can see that near temperature 
$T\approx 0.90$ there is a second order phase transition.
The change of the slope of the curves and the fluctuations in the simulations
near $T\approx 0.9$ is consistent with the existence of a second order
phase transition.

\begin{figure}[t] %  figure placement: here, top, bottom, or page
   \centering
   \includegraphics[width=2.9in]{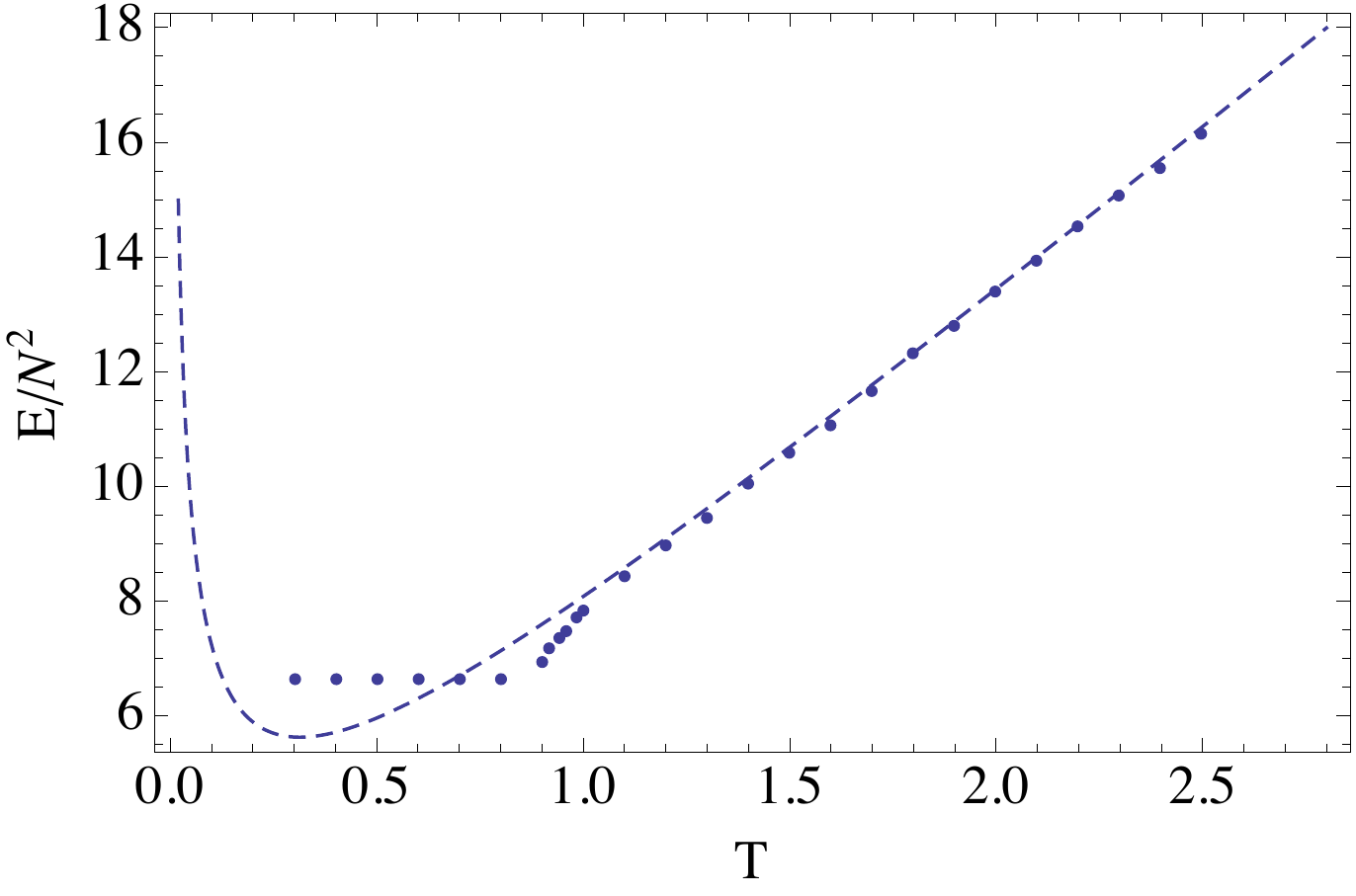} 
      \includegraphics[width=2.9in]{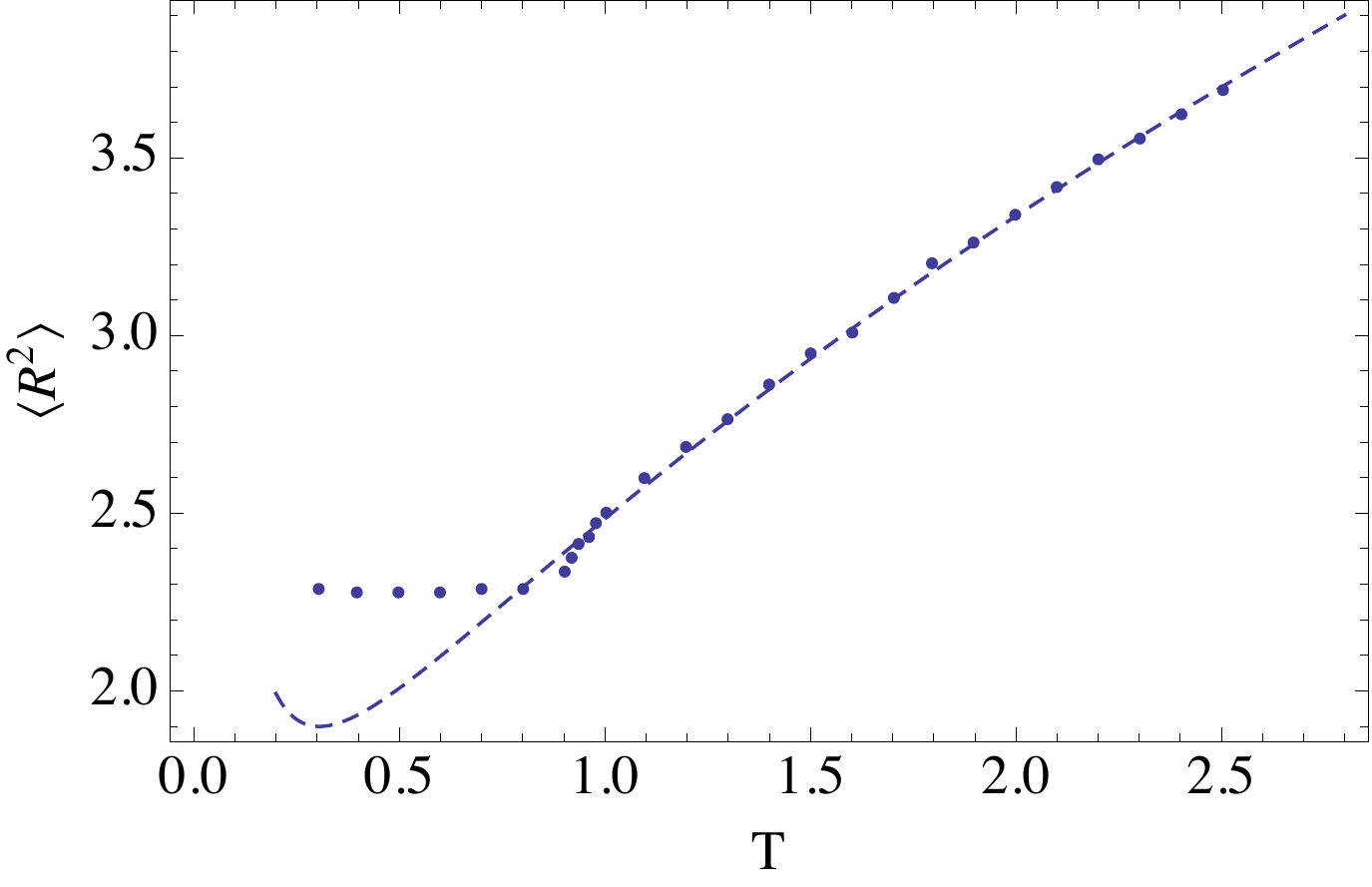} 
 \caption{\small Plots of the scaled energy $E/N^2$ and the ``extent of space'' $\langle R^2\rangle$ as functions of the temperature. The dashed curves correspond to the high temperature behaviour obtained in~\cite{Kawahara:2007ib}. One can see that near $T\approx 0.9$ the plots suggest the existence of a second order phase transition. The energy and temperature in the plots are in units of $\lambda^{1/3}$.}
   \label{fig:2}
\end{figure}

In figure \ref{fig:2} we present plots of the energy and ``extent of
space'' as functions of the temperature, for $N=16$ and lattice
spacing $a\approx 0.05$. The dashed
curves represent the analytical high temperature behaviour obtained in~\cite{Kawahara:2007ib}. Our results agree very well with the
corresponding studies in~\cite{Kawahara:2007fn, Mandal:2011hb, Azuma:2014cfa}.

\begin{figure}[t] %  figure placement: here, top, bottom, or page
   \centering
   \includegraphics[width=3.9in]{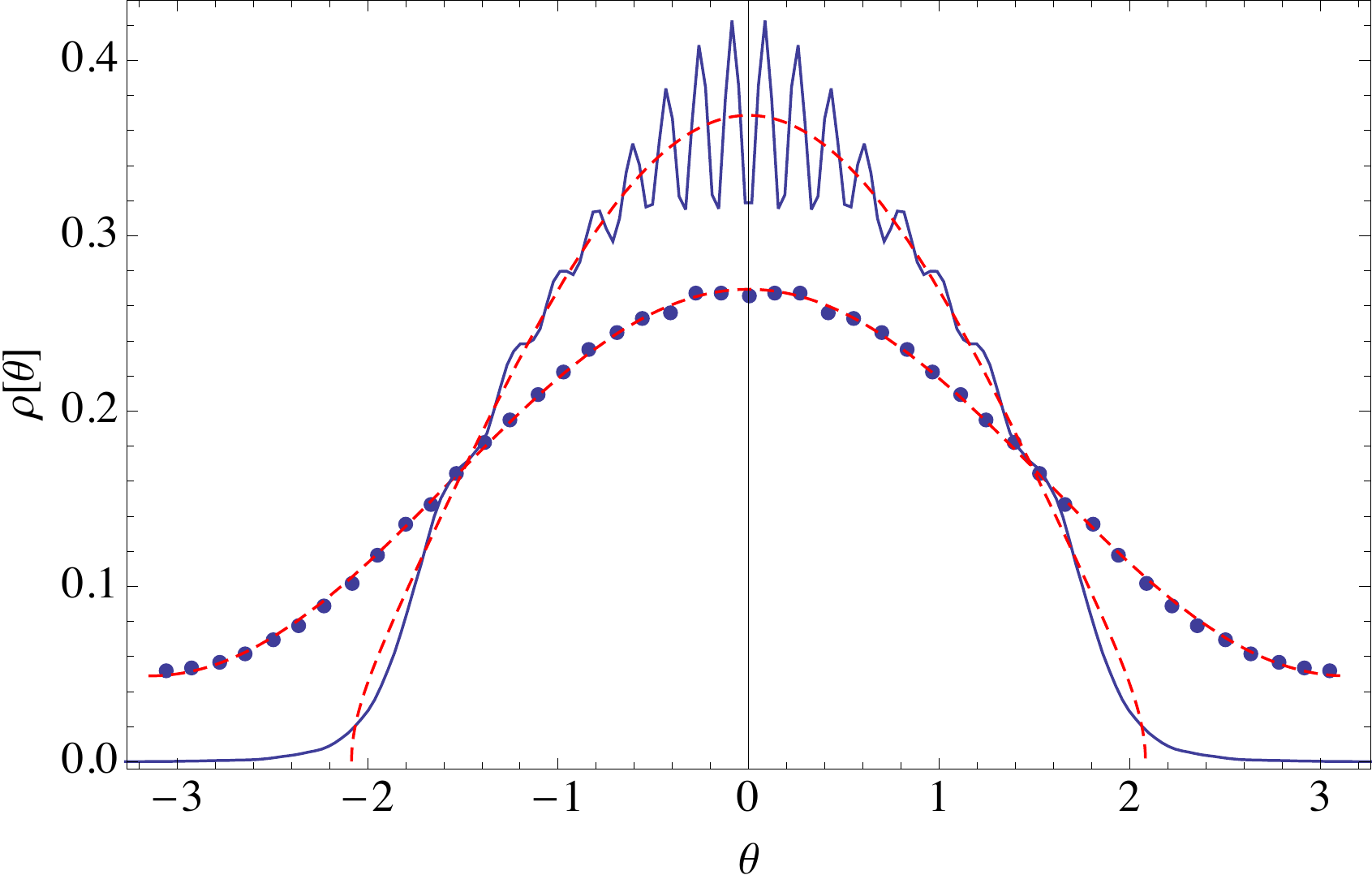} 
 \caption{\small  Plots of the distribution of the holonomy $P$ for temperatures $T=0.900$ (the confined or ungapped) and $T=.9006$ (the deconfinied or gapped). The plots are for size $N=16$ and lattice spacing $a\approx 0.05$. The dashed curves correspond to fits to the Gross-Witten gapped and ungapped distributions.}
   \label{fig:3}
\end{figure}

A more detailed analysis of the temperature range close to the
transition revealed that there are in fact two transitions. To uncover
more detail on the nature of the phase transition the authors
of~\cite{Kawahara:2007fn} analysed the distribution of the holonomy
matrix near the phase transition and uncovered behaviour consistent
with the Gross-Witten model \cite{Gross:1980he} and concluded that the
holonomy undergoes a transition from a uniform distribution at
$T_{c2}\simeq 0.8758(9)$ to a gapped distribution at
$T_{c2}=0.905(2)$.
In figure \ref{fig:3} we present our plots of the distribution of the
holonomy around the phase transition. The dashed curves in the plot
represent fits with the gapped and ungapped forms of the Gross-Witten
distribution which are in excellent agreement with those of
\cite{Kawahara:2007fn, Mandal:2011hb, Azuma:2014cfa}. We have not attempted to refine their
results, rather our purpose is to uncover a hidden Gaussian structure
in the model.

\subsection{Gap and eigenvalue distribution}

In this section we investigate the eigenvalue distribution of the
scalar fields. We also perform studies of the mass of the theory at
zero temperature. Our results suggest that at all temperatures the
eigenvalue distribution of any one of the $X^i$ is given by a Wigner
semicircle, with a radius $R_\lambda$ following the temperature
behaviour of the observable $\langle R^2\rangle$.\footnote{The
  semicircle law implies $R_\lambda^2 =\frac{4}{p}\langle R^2\rangle$.}
Therefore, we conclude that while the radius of the distribution
experiences a phase transition the shape of the distribution remains
unchanged. We believe that the main reason for this behaviour is that
for nine scalar fields the commutator squared term can be replaced and
approximated by a quadratic mass term in the spirit
of~\cite{Hotta:1998en}, where an expansion at large number of scalar
fields has been developed. Generalising these techniques, we are able to
obtain an estimate of the mass, which agrees very well with both the
gap measured from correlation functions and the radius of the
distribution which are obtained from Monte Carlo simulations.

In the limit of high temperature the model reduces
to the 10-dimensional Yang-Mills matrix model 
considered in~\cite{Hotta:1998en}. The obtained
behaviour of the radius is in agreement with the large temperature
expansion performed in~\cite{Kawahara:2007ib} and provides an
analytic approximation to the leading coefficients in this expansion.

We begin by considering the model at zero temperature. In this case
the holonomy can be completely gauged away and the model
simplifies. Furthermore, at zero temperature the correlator:
\begin{equation}
\left\langle{\rm Tr}\left(X^1(0)X^1(t)\right)\right\rangle\propto e^{-m\,t} +\dots\ ,
\end{equation}
captures the gap $\Delta m = E_1-E_0$ of the theory. To calculate the gap in the discrete theory, we periodically identify the time direction with period $\beta$:
\begin{equation}\label{corel}
\left\langle{\rm Tr}\left(X^1(0)X^1(t)\right)\right\rangle =A\,( e^{-m\,t} +e^{-m(\beta -t)}) \ ,
\end{equation}
Note that although formally $\beta$ is the same parameter that we have
at finite temperature, since we set the holonomy to zero here its
meaning is just a periodic coordinate as opposed to inverse temperature. Our
result for the correlator for $N=30$, $\beta=10$ and lattice spacing
$a=0.25$ is presented on the left in figure \ref{fig:4}. The fitting
curve is given by equation (\ref{corel}) and when we perform a two parameter
fit we obtain $A\approx 7.50\pm 0.2$ and
$m\approx (1.90\pm.01)\,\lambda^{1/3}\,$. However, for Gaussian scalar
fields of mass $m$ we have $A=\frac{N}{2m(1-{\rm e}^{-\beta m})}$ and
performing a one parameter fit for $m$ yields $m=1.965\pm0.007$
and $A=7.63\pm 0.03$ On the
right we have presented a plot of the eigenvalue distribution of one
of the matrices for the same parameters. The fitting curve represents
a Wigner semicircle of radius $R_\lambda\approx 1.01$. The fact that the
theory is gapped and that the eigenvalue distribution is a semicircle
suggests that that at low temperate the model has an effective action:
\begin{figure}[t] %  figure placement: here, top, bottom, or page
   \centering
   \includegraphics[width=2.9in]{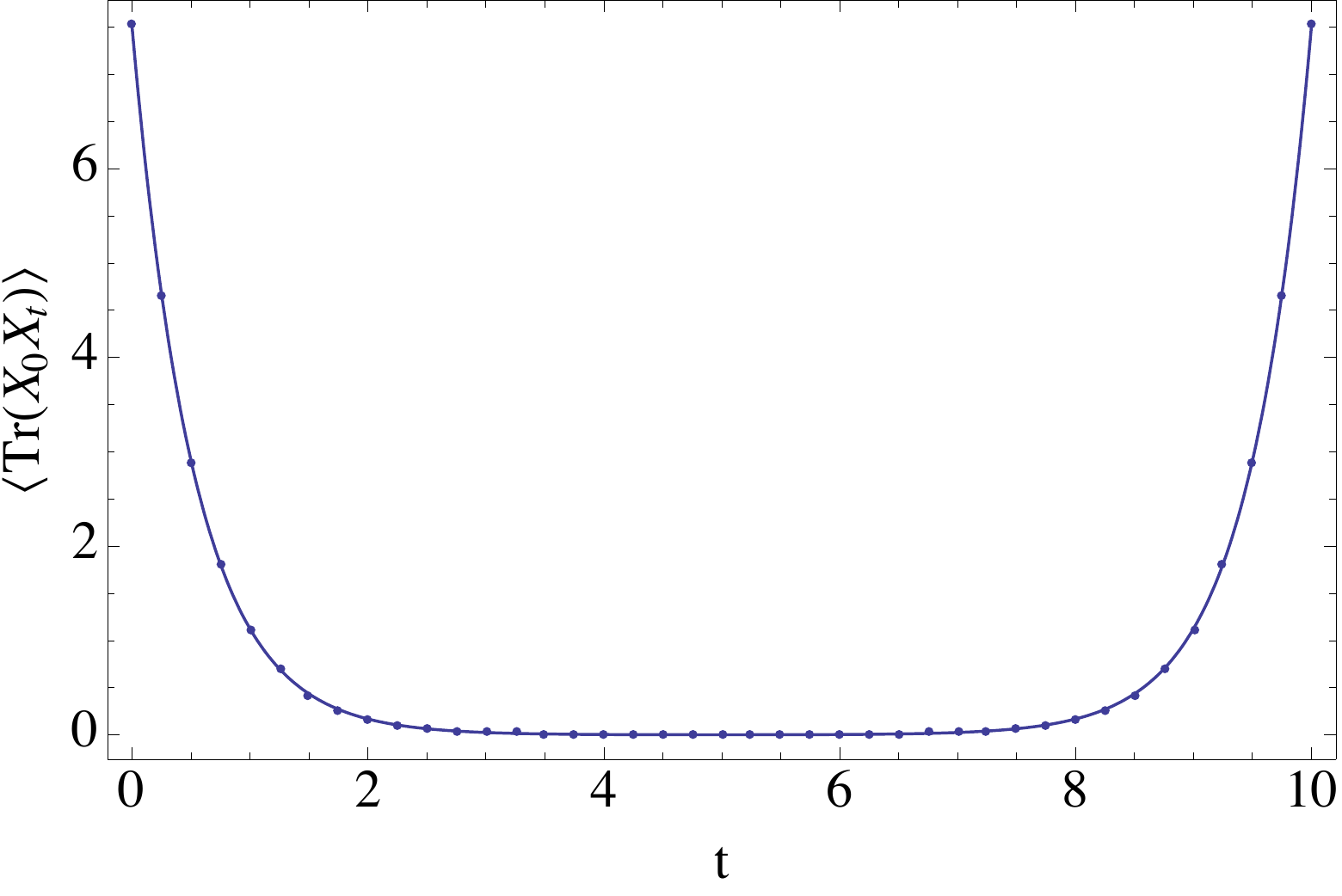} 
   \includegraphics[width=2.9in]{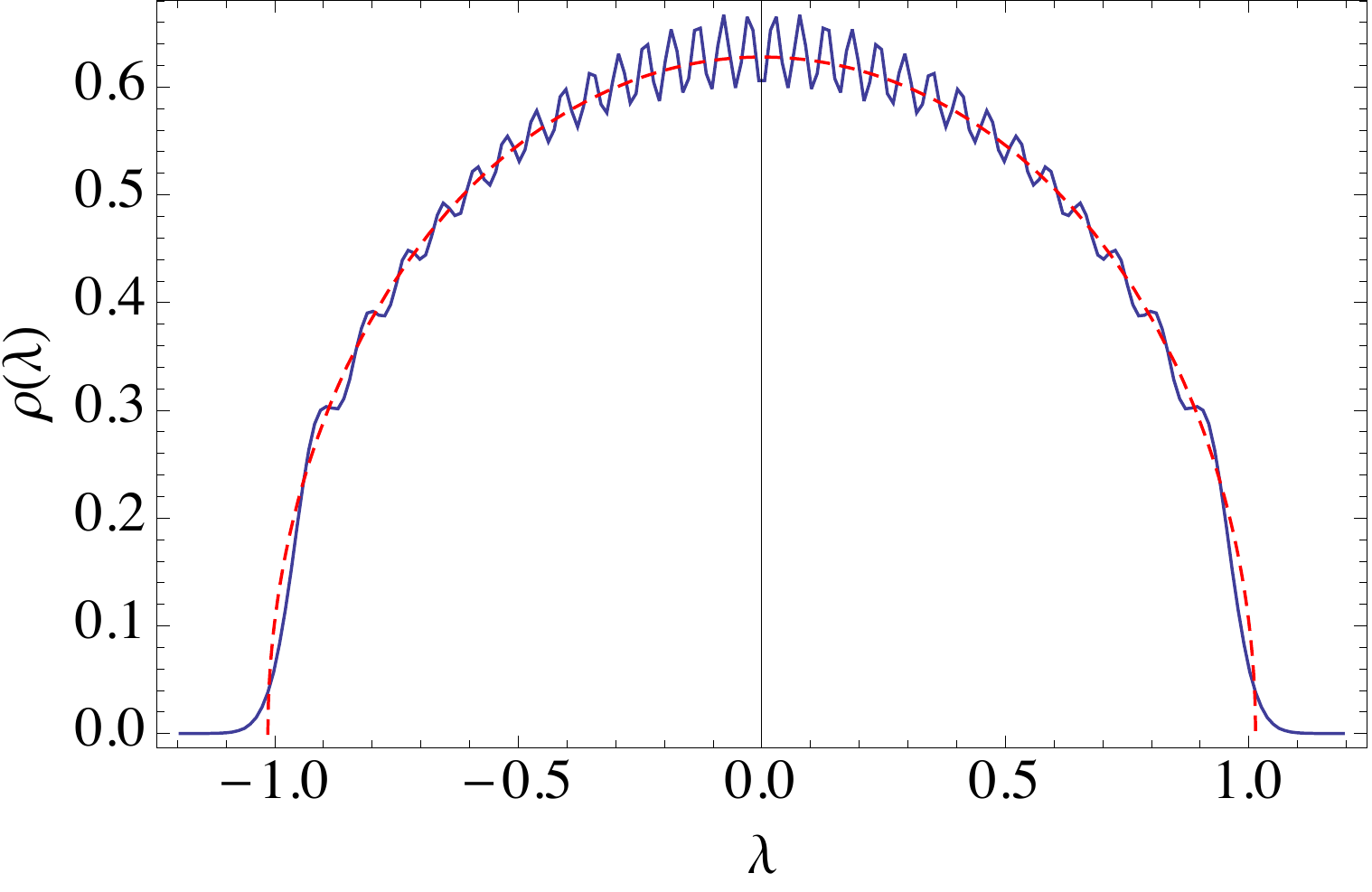} 
 \caption{\small {\it On the left:} A plot of the correlator
   $\left\langle{\rm Tr}\left(X^1(0)X^1(t)\right)\right\rangle$ for
   $N=30$, $\beta =10$ and lattice spacing $a=0.25$. The fitting curve
   is given by equation (\ref{corel}) with $A=\frac{N}{2m(1-{\rm e}^{-\beta m})}$
   and with parameters $m \approx (1.965\pm.007)\lambda^{1/3}\,$. {\it On
     the right:} A plot of the eigenvalue distribution of one of the
   scalars for the same parameters. The fitting curve represent a
   Wigner-semicircle of radius $R_\lambda\approx 1.01\,$. }
   \label{fig:4}
\end{figure}
\begin{equation}
S_{\rm eff}=N\int\limits_{-\infty}^\infty dt\,{\rm Tr}\left( \frac{1}{2}(\dot X^i)^2+\frac{1}{2}m^2 (X^i)^2\right)
\label{GaussianModel}
\end{equation}
for each of the matrices $X^i$. It is well known \cite{Brezin:1977sv}
that for the action (\ref{GaussianModel}) the eigenvalue distribution
of $X$ is given by a Wigner semicircle of radius:
\begin{equation}
R_\lambda =\sqrt{\frac{2}{m}}\approx 1.009\pm.002,
\end{equation}
where we have substituted $m \approx 1.965\pm .007$.
This agrees nicely (within errors) with the result for $R_\lambda\approx 1.01$
obtained by fitting the actual distribution. It is also in excellent
agreement with the large $p$ theoretical prediction of \cite{Mandal:2011hb},
\begin{equation}
R_\lambda(p)=\sqrt{\frac{2}{p^{1/3}}}(1+\frac{1}{p}(\frac{7\sqrt{5}}{30}-\frac{9}{32})+\cdots)\simeq 1.0068
\end{equation}
\subsection{$1/D$ expansion}
Adapting the techniques developed in~\cite{Hotta:1998en} to the time
dependent case that we are considering we can obtain a theoretical
estimate of the radius at low temperature\footnote{The next order corrections for the current model were computed in \cite{Mandal:2011hb} but we were unaware of this work at the time of writing.}. Let us consider again the
action of the bosonic model (\ref{BosS}):
\begin{equation}\label{BosS1}
S_b = N\,\int_0^{\beta}dt\,{\rm tr}\left\{\frac{1}{2}({\cal D}_t {X^i})^2-\frac{1}{4}[X^i,X^j]^2\right\}\ ,
\end{equation}
where we have rescaled so that $\beta=\frac{\lambda^{1/3}}{T}$
(effectively we set $g=N^{-1/2}$). The commutator square term can be
written as:
\begin{equation}
{\rm Tr}[X^i,X^j]^2={\rm Tr}\left([t^a,t^c][t^b,t^d]\right)X^i_aX^i_b\,X^j_cX^j_d=\lambda^{abcd}X^i_aX^i_b\,X^j_cX^j_d\ ,
\end{equation}
where $t^a$ are $SU(N)$ generators normalised to ${\rm Tr\,}t^at^b =\delta^{ab}$ and the tensor $\lambda^{abcd}$ is given by:
\begin{equation}
\lambda^{abcd}=\frac{1}{2}{\rm Tr}\left([t^a,t^c][t^b,t^d]+[t^a,t^d][t^b,t^c]\right)\ ,
\end{equation}
It is convenient also to define the inverse kernel of $\lambda^{abcd}$
satisfying:
\begin{equation}
\mu_{abcd}\,\lambda^{cdef} =\delta_a^e\delta_b^f\ ,~~~\lambda^{abcd}\,\mu_{cdef} =\delta^a_e\delta^b_f\ .
\end{equation}
We will also use the identities:
\begin{equation}
\lambda^{abcd}\,\delta_{cd} =-2N\,\delta^{ab}\ ,~~~~\mu_{abcd}\,\delta^{cd} =-\frac{1}{2N}\,\delta_{ab}\ .
\end{equation}
The action (\ref{BosS1}) can then be written as:
\begin{equation}
S_b=N\,\int_0^{\beta}dt\,{\rm Tr}\left(\frac{1}{2}({\cal D}_t {X^i})^2\right)-\frac{N}{4}\lambda^{abcd}\int_0^{\beta}dt\,X^i_aX^i_b\,X^j_cX^j_d\ .
\end{equation}
Our next step is to add to the action the term\footnote{Note that we can always add this term since $\int {\cal D}k\,e^{-\Delta S}={\rm const}$ }:
\begin{equation}
\Delta S =\frac{N}{4}\mu_{abcd}\int\limits_0^\beta dt\left(k^{ab}+\lambda^{abef}X^i_eX^i_f\right)\left(k^{cd}+\lambda^{cdgh}X^i_gX^i_h\right)\ ,
\end{equation}
the action $S_b'=S_b+\Delta S$ then becomes:
\begin{equation}
S_b'=\frac{N}{2}\,\int_0^{\beta}dt\,\left\{{\rm Tr}({\cal D}_t {X^i})^2+k^{ab}X_a^iX_b^i\right\}+\frac{N}{4}\mu_{abcd}\int\limits_0^\beta\,dt\, k^{ab}k^{cd}\ .
\label{ActionSb'}
\end{equation}
Next we define:
\begin{equation}
k_{\,ij,lm}\equiv t_{ij}^a \,k^{ab}\,t_{lm}^b\ ,
\end{equation}
From the definition of $k_{ij,lm}$ it follows that it is traceless
with respect to the first and the second pair of indices and we can
invert: $k^{ab}=t^a_{ij}\,k_{ji,ml}\,t^b_{lm}$. Substituting in the
action (\ref{ActionSb'}), Fourier transforming (via $X = \sum_n e^{i
  \frac{2\pi n}{\beta}t}\tilde X_n$) and assuming $k_{ab}$ is time
independent we obtain:
\begin{equation}
S_b'=\frac{N}{2}\sum\limits_{n=-\infty}^\infty \tilde X_{-n,\, ij}\left(\left(\frac{2\pi \,n +\theta_i-\theta_j}{\beta}\right)^2P_{jl,ml}+k_{jl,ml}\right)\tilde X_{n,\,lm}+\frac{\beta\,N}{4}\mu_{abcd}\, k^{ab}k^{cd}\ ,
\end{equation}
where we have defined the projector on traceless matrices:
\begin{equation}
P_{ij,lm}=t_{ij}^a\,t^a_{lm}=\delta_{im}\delta_{jl}-\frac{1}{N}\delta_{ij}\delta_{lm}\ 
\end{equation}
and assumed that $k$ is constant. Defining also the double index matrix :
\begin{equation}\label{matW}
W(n)_{ij,lm}\equiv \left(\frac{2\pi \,n +\theta_i-\theta_j}{\beta}\right)^2P_{jl,ml}+k_{jl,ml}\ ,
\end{equation}
we can integrate out the X's:
\begin{equation}
\int {\cal D}X\,{\cal D}k\,e^{-S_b'}\propto\int {\cal D}k \,\prod_n{\rm Det}^{-{p}/{2}}\left(P.W(n).P\right)e^{-\frac{\beta\,N}{4}\mu_{abcd}\, k^{ab}k^{cd}}\ .
\end{equation}
The effective action for the field $k$ then becomes:
\begin{equation}
S_{\rm eff}[k]=\frac{p}{2}\sum\limits_n{\rm Tr}\log\left(P.W(n).P\right)+\frac{\beta\,N}{4}\mu_{abcd}\, k^{ab}k^{cd}\ .
\end{equation}
We now notice that the first term in the expression for the matrix $W$ (\ref{matW}) commutes with the projector $P$. It is natural then to consider an ansatz for $k$ which also have that property. Thus we consider:
\begin{equation}
k_{ij,lm}\equiv k_{ij}\,P_{ij,lm}= P_{ij,lm}\,k_{lm} \,
\end{equation}
The last equality is possible only if all diagonal components of $k_{ij}$ are the same (namely $k_{ii}=k_{jj}$ for all $i,j$) we also choose $k_{ij}$ to be symmetric. Then:
\begin{eqnarray}
&&W(n)_{ij,lm}=\Delta_{ij}(n)\,P_{ij,lm} = \,P_{ij,lm}\Delta_{lm}(n) \ ,\\
&&\Delta_{ij}(n)\equiv \left(\frac{2\pi \,n +\theta_i-\theta_j}{\beta}\right)^2+k_{ij} \nonumber\ ,
\end{eqnarray}
and we have for all powers $r$ that $(P.W(n).P)_{ij,lm}^r
=P_{ij,lm}\,(\Delta_{ij}(n))^r$. Therefore,
\begin{equation}
S_{\rm eff}[k]=\frac{p}{2}\sum_n\sum_{ij} P_{ij,ji}\log(\Delta(n)_{ij})+\frac{\beta\,N}{4}\mu_{abcd}\, k^{ab}k^{cd}\ .
\end{equation}
The corresponding saddle point equation $S_{\rm eff}'[k]=0$ becomes:
\begin{equation}
\frac{p}{2}\sum_n\frac{P_{ij,ji}}{\left(\frac{2\pi \,n +\theta_i-\theta_j}{\beta}\right)^2+k_{ij}  }+\frac{\beta\,N}{2}\mu_{abcd}\, k^{ab}t^c_{ij}t^d_{ji} =0\ .
\end{equation}
We can now sum the series to obtain:
\begin{equation}\label{saddlek}
\frac{p}{\sqrt{k_{ij}}}\,\frac{\sinh(\beta\sqrt{k_{ij}})}{\cosh(\beta\sqrt{k_{ij}})-\cos(\theta_i-\theta_j)}P_{ij,ji}+{2N}\mu_{abcd}\, k^{ab}t^c_{ij}t^d_{ji} =0\ .
\end{equation}
In principle we can solve for $k_{ij}$ in terms of
$\theta_i-\theta_j$, but we will restrict ourselves to extract the low
temperature dependence. The first term in equation (\ref{saddlek}) has
the following expansion:
\begin{equation}
\frac{\sinh(\beta\sqrt{k_{ij}})}{\cosh(\beta\sqrt{k_{ij}})-\cos(\theta_i-\theta_j)}=1+2\cos(\theta_i-\theta_j)\,e^{-\beta\sqrt{k_{ij}}}+O(e^{-2\,\beta\sqrt{k_{ij}}})
\end{equation}
One can see that the effect of the holonomy is exponentially
suppressed at low temperature, To leading order we can then consider a
symmetric ansatz $k_{ij}=k_0$, which also implies
$k^{ab}=k_0\,\delta^{ab}$. We obtain:
\begin{equation}
\frac{p}{\sqrt{k_0}}P_{ij,ji}-k_0\,P_{ij,ji} =0~~\therefore k_0 =p^{2/3}\ .
\end{equation}
Substituting into the action (\ref{ActionSb'}) to leading order we obtain:
\begin{equation}
S_b=N\int\limits_{-\infty}^\infty dt\,{\rm Tr}\left(\frac{1}{2}(\dot X^i)^2+\frac{p^{2/3}}{2}(X^i)^2\right)
\end{equation}
and the corresponding radius of the eigenvalue distribution is:
\begin{equation}
R_0 =\left(\frac{8}{p}\right)^{1/6}\approx 0.98\ .
\end{equation}
This result agrees within a few percent with $R_{\lambda}\approx 1.01$ obtained by fitting the distribution in figure \ref{fig:4}. 

\begin{figure}[t] %  figure placement: here, top, bottom, or page
   \centering
   \includegraphics[width=2.9in]{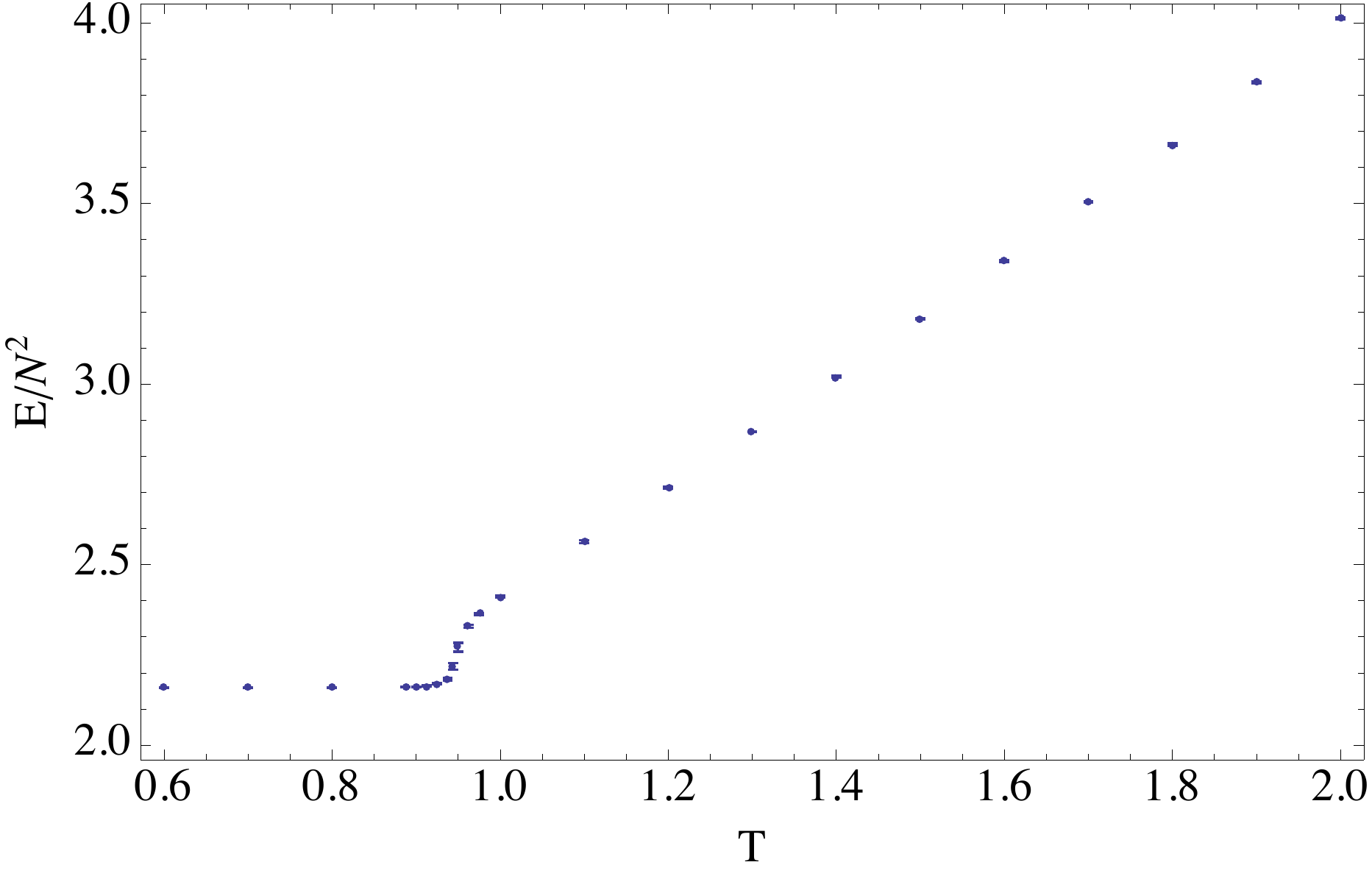} 
   \includegraphics[width=2.9in]{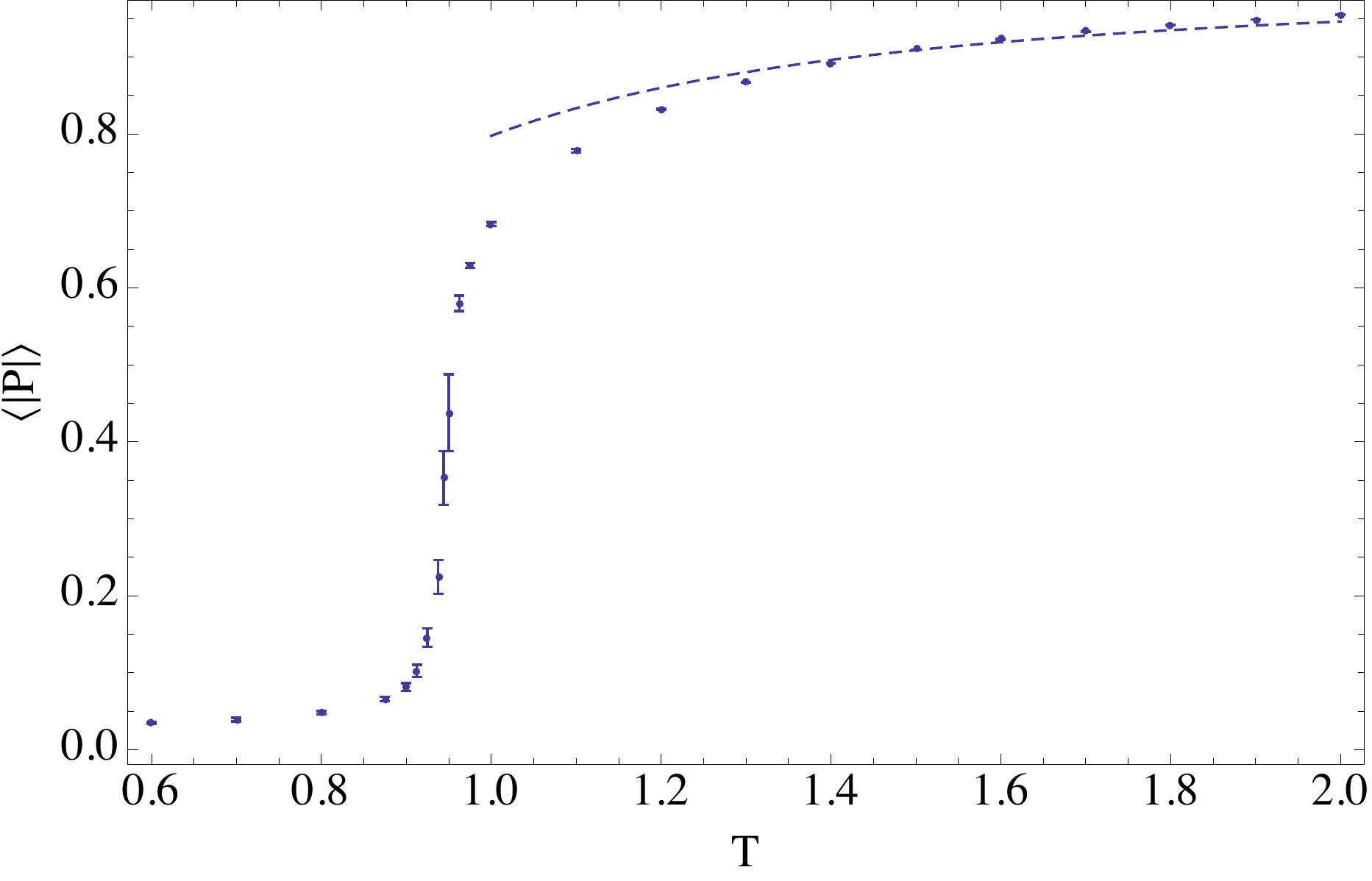} 
 \caption{\small {\it On the left:} A plot of the energy of the gauged
   gaussian model for $N=32$ lattice spacing $a\approx0.05$. The red
   curve represents Wigner semicircle. {\it On the right:} A plot of
   the expectation value of the Polyakov loop for $N=32$ lattice
   spacing $a\approx0.05$. The second order phase transition takes place at
   $T\approx 0.95$. The energy and temperature in the plots are in
   units of $\lambda^{1/3}$.}
   \label{fig:4A}
\end{figure}

The exponential suppression of the holonomy corrections to the low
temperature saddle point for $k^{ab}$ suggest that a lot of the
physics of the model (at least at low temperature) can be captured by
the action:
\begin{equation}
S_b=N\int\limits_{0}^\beta dt\,{\rm Tr}\left(\frac{1}{2}({\cal D}_t X^i)^2+\frac{p^{2/3}}{2}(X^i)^2\right)\ ,
\label{gaugedgaussian}
\end{equation}
where we restored the gauge field in the covariant
derivatives. Surprisingly this model knows lots about the phase
transition of the full model. An analysis similar to that above shows
that the critical temperature for $p$ adjoint gauged Gaussian matrices
as in \cite{Mandal:2009vz} with mass $m$
occurs at $T_{c}^{Gaussian}=\frac{m}{\ln p}$ which for $m=p^{1/3}$ yields
$Tc=0.9467$.  In figure \ref{fig:4A} we have presented
our results for the energy and the Polyakov loop. Note that in this
approximation the scaled energy $E/N^2$ is equal to the extent of
space $\langle R^2\rangle$.
The plots are for $N=32$ and lattice spacing $a\approx 0.05$. One can
see that both the energy and the Polyakov loop exhibit the same
behaviour as for the full model. There again appear to be two distinct
transitions with the higher temperature one taking place at $T\approx
.95$ and again appearing to be second order. It is slightly shifted
towards high temperatures relative to the critical temperature for the
full model. If instead of mass $m =p^{1/3}\lambda^{1/3}$ we use the
value $m\simeq (1.90\pm.01)\lambda^{1/3}$ the phase transition is
shifted in the opposite direction (just bellow $T=0.9$). This
indicates that if one fits the mass parameter one can improve even
further the agreement of the gauged gaussian model and the full
one. The dashed curve in the second plot is the theoretical prediction
for the high-temperature behaviour of the Polyakov loop, again one can
observe a very good agreement. The high temperature behaviour of the
energy on the other side disagrees with the corresponding behaviour of
the full model. This is not surprising since we derived the effective
action at low temperature and the dominant behaviour at high
temperature is dominated by the highest power of the potential which
has been changed from quartic to quadratic. 
\begin{figure}[t] %  figure placement: here, top, bottom, or page
   \centering
   \includegraphics[width=3.6in]{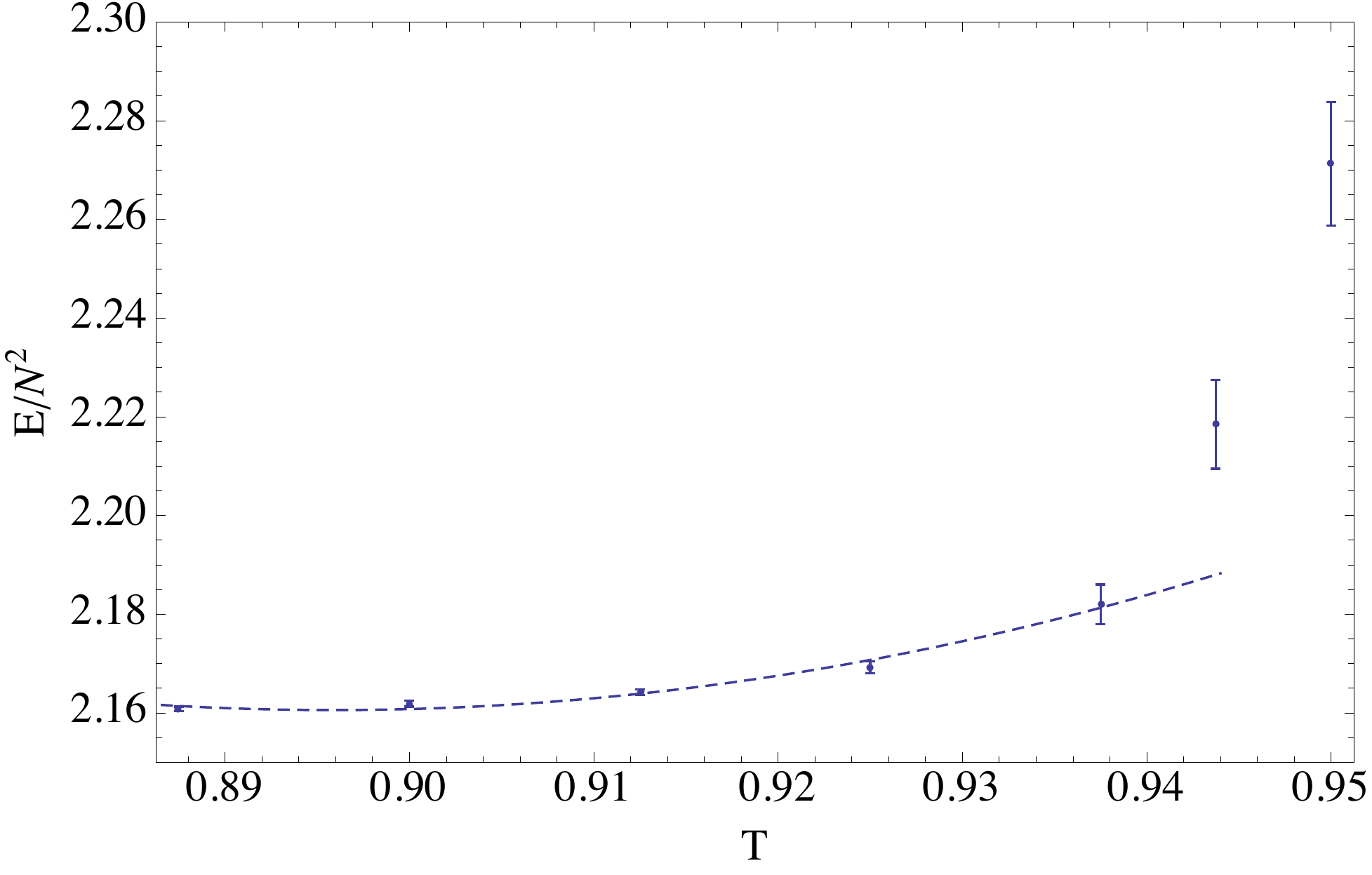} 
 \caption{\small  A zoomed in plot of the energy versus the temperature. The dashed curve represents a fit to equation (\ref{eq-fit-par}) with $T_0\approx 0.896$.}
   \label{fig:4B}
\end{figure}
One can also see that at low temperature the energy remains constant. 

In figure \ref{fig:4B} we have presented a plot of the energy versus the temperature zoomed in near the phase transition. The dashed curve represents a fit with:
\begin{equation}\label{eq-fit-par}
E-\epsilon_0 = C (T-T_0)^2
\end{equation}
with parameter $T_0\approx 0.896$. This indicates that the third order phase transition that the full model exhibits \cite{Kawahara:2007fn, Mandal:2011hb} is also captured by the gauged gaussian model.

One can perform the large $p$ analysis (see \cite{Hotta:1998en}) in
the high temperature limit where the model now has $p+1$ matrices (the
holonomy becomes the additional matrix) and predict that the model in
this limit again becomes Gaussian but now the field $k_{ab}$ includes
the holonomy and the saddle becomes $k_{ab}=\sqrt{p+1}\delta_{ab}$
which again predicts a Gaussian matrix model with a high temperature
effective action $S_{eff}=\frac{\sqrt{p+1}}{2}Tr((X^i)^2)$ and
consequently a Wigner semicircle distribution for the eigenvalues os
$X^i$.

\begin{figure}[t] %  figure placement: here, top, bottom, or page
   \centering
   \includegraphics[width=2.9in]{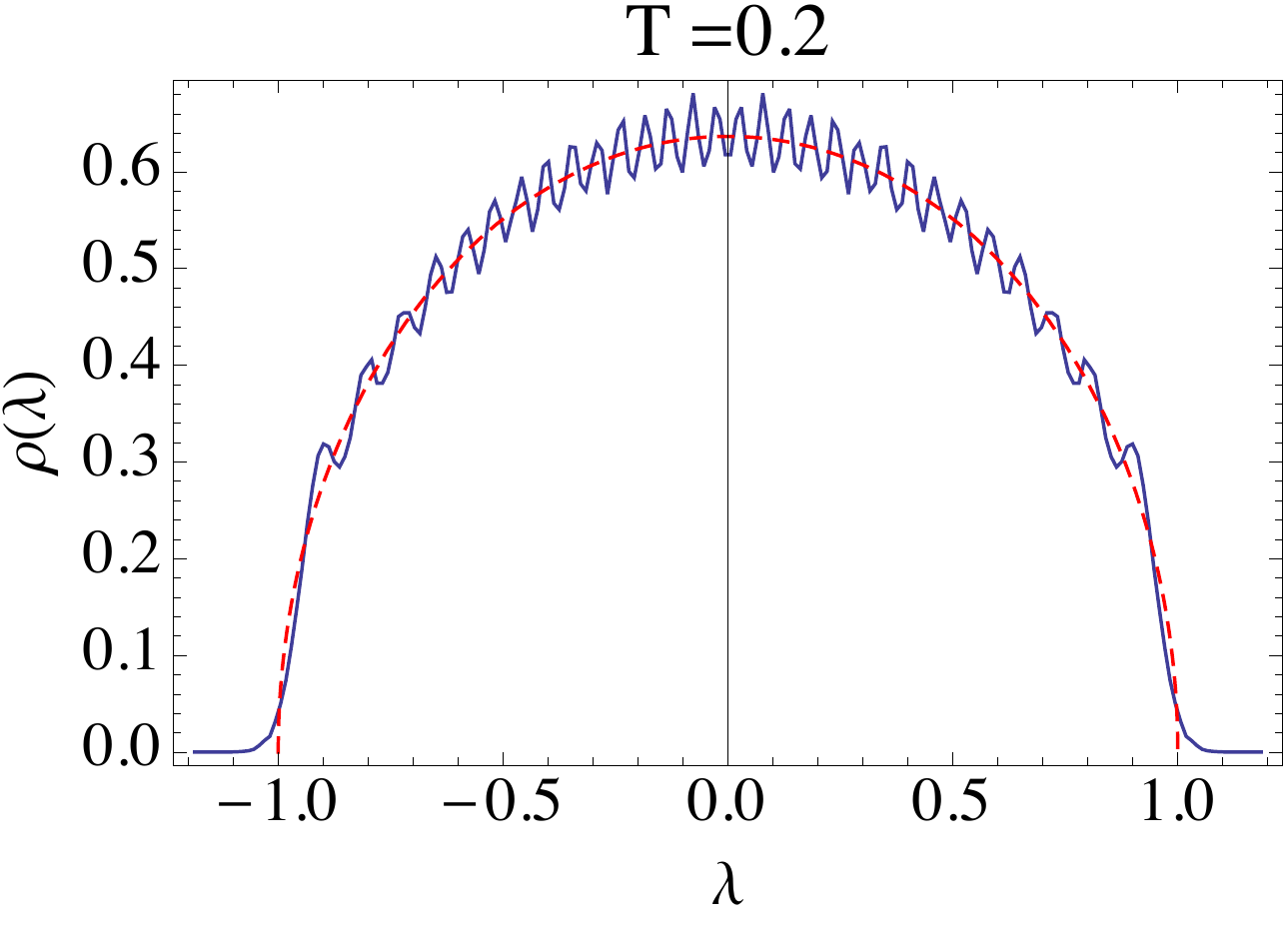} 
   \includegraphics[width=2.9in]{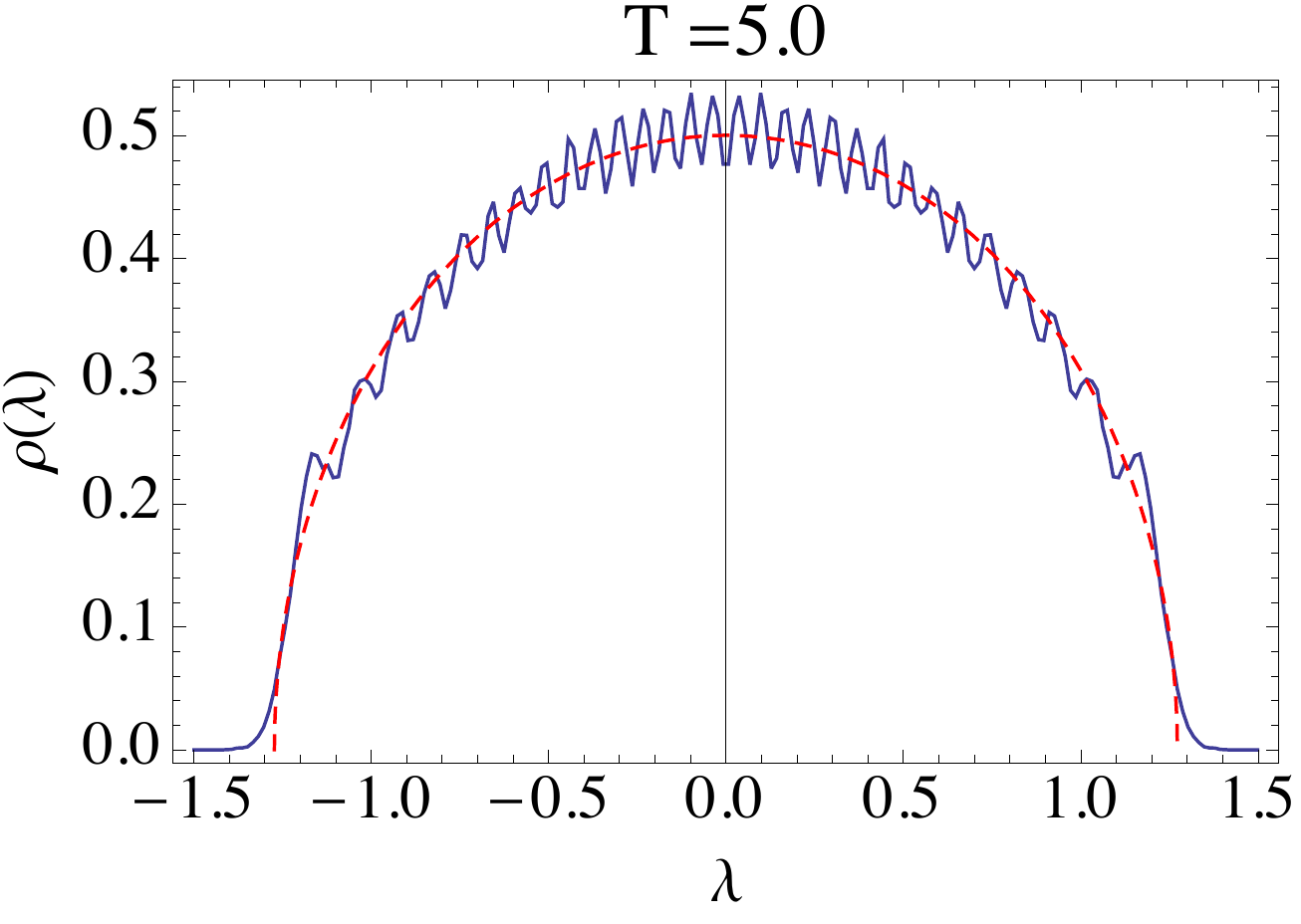} 
 \caption{\small {\it On the left:} A plot of the eigenvalue
   distribution for $N=30$, $T=.2$ and lattice spacing $a=.25$. The
   red curve represents Wigner semicircle. {\it On the right:} A plot
   of the eigenvalue distribution for $N=30$, $T=5.0$ and lattice
   spacing $a=.05$. Again the red curve represents Wigner semicircle.}
   \label{fig:5}
\end{figure}

We conclude this section by presenting results for the eigenvalue
distribution of the gauged gaussian model at finite temperature. In
figure \ref{fig:5} we presented plots of the distribution for $T=0.2$,
$N=30$ (left) and $T=5.0$, $N=30$ (right). The red dashed curves show
a Wigner semicircle. One can see that the shape of the eigenvalue
distribution does not change with temperature.

%
%%%%%%%%%%%%%%%%%%%%%%%%%%%%%%%%
%%%%%%%%%%%%%%%%%%%%%%%%%%%%%%%%
%%%% BEGINNING OF SUSY
%%%%%%%%%%%%%%%%%%%%%%%%%%%%%%%%
%%%%%%%%%%%%%%%%%%%%%%%%%%%%%%%%

%\newpage

\section {The supersymmetric model on the lattice }
\label{supersymmetric section}
In this section we consider the full supersymmetric BFSS model on the
lattice. The model has been simulated using non-lattice approach
in~\cite{Anagnostopoulos:2007fw} and using lattice discretisation
in~\cite{Catterall:2008yz} and \cite{Kadoh:2015mka}. The main goal of
these studies has been to compare the low temperature regime of the
model to the holographically dual black hole geometry. The authors of
refs. \cite{Anagnostopoulos:2007fw} and \cite{Kadoh:2015mka} also
compared the high temperature regime of the model with the high
temperature expansion performed in~\cite{Kawahara:2007ib}. Our goal is
to reproduce some of these studies and calibrate our code.

A naive discretisation of the action (\ref{BFSS-9D}) would result in a fermion doubling. This can be avoided ~\cite{Catterall:2008yz} if the charge conjugation matrix is taken to be $C_9=1_8\otimes \sigma_1$. \footnote{It is analogous to using staggered fermions, which in one dimension complete removes the doublers.} Constructing a basis for which $C_9$ is of this form is relatively straightforward. For example one can tensor up the Majorana basis in seven euclidean dimensions $\tilde\gamma_E^a$:
\begin{eqnarray}\label{9gammas}
\gamma^a &=& -\tilde \gamma_E^a \otimes \sigma_3\ , ~~~{\rm for}~a=1,\dots,7\ ,\nonumber \\
\gamma^8 &=& 1_8\otimes\sigma_2\ ,\nonumber \\
\gamma^9 &=& 1_8\otimes\sigma_1\ ,
\end{eqnarray}
and verify that indeed $C_9$ is of the desired form (it also satisfies $C_9=\gamma^9$). We next proceed in discretising the action (\ref{BFSS-9D}). Since the bosonic part of the action is identical to the one considered in section \ref{bosonic section} , we will consider only the fermionic part of the action:
\begin{equation}
S_f = \frac{1}{2g^2}\int d\tau \,{\rm tr}\left\{\psi^{\alpha} C_{9\,\alpha\beta}\,{\cal D}_\tau\psi^{\beta} -\psi^{\alpha} (C_{9}\gamma^i)_{\alpha\beta}\,[X^i,\psi^{\beta}]\right\}\ .
\end{equation}
We begin by splitting the fermions into two eight component fermions: $\psi=(\psi_1,\psi_2)$ and defining the forward and backward derivatives $D_\pm$:
\begin{eqnarray}
({\cal D}_-W)_n&=&(W_n-U_{n,n-1}W_{n-1}U_{n-1,n})/{a} \ , \nonumber \\
({\cal D}_+W)_n&=&(U_{n,n+1}W_{n+1}U_{n+1,n}-W_n)/{a}\ .
\end{eqnarray}
One can show that the discretised kinetic term then becomes:
\begin{eqnarray}
S_f^{\rm kin}&=&\frac{1}{2g^2}\int d\tau\,{\rm tr}\left(\psi^{\alpha} C_{9\,\alpha\beta}\,{\cal D}_\tau\psi^{\beta}\right) = \frac{a}{2g^2}\sum_{n=0}^{\Lambda-1}{\rm tr}\left\{\psi_{1,n}^T({\cal D}_-\psi_2)_n+\psi_{2,n}^T({\cal D}_+\psi_1)_n    \right\} =\\
&&=\frac{1}{g^2}{\rm tr}\left\{-\sum_{n=0}^{\Lambda-1}\psi^T_{2,n}\psi_{1,n}+\sum_{n=0}^{\Lambda-2}\psi^T_{2,n}U_{n,n+1}\psi_{1,n+1}U_{n+1,n} \pm \psi^T_{2,\Lambda-1}U_{\Lambda-1,0}\psi_{1,0}U_{0,\Lambda-1}\right\}\,\,\nonumber\ ,
\end{eqnarray}
where the plus/minus sign in the last term corresponds to periodic/anti-periodic boundary conditions for the fermions.\footnote{Namely the conditions $\psi_{-1}=\pm\psi_{\Lambda-1}$ and  $\psi_{\Lambda}=\pm\psi_{0}$. }
Using the gauge from the previous subsection when the holonomy is concentrated on a singe link we can write $S_f^{\rm kin}$ as:
\begin{equation}
S_f^{\rm kin}=\frac{1}{g^2}{\rm tr}\left\{-\sum_{n=0}^{\Lambda-1}\psi^T_{2,n}\psi_{1,n}+\sum_{n=0}^{\Lambda-2}\psi^T_{2,n}\psi_{1,n+1}\pm \psi^T_{2,\Lambda-1}D\,\psi_{1,0}\,D^{\dagger}\right\}\ .
\end{equation}
Since all fields transform in the adjoint of $SU(N)$ instead of
dealing with matrices we can use the corresponding real components:
$X^a={\rm tr} (t^aX)$ and $\psi^a = {\rm tr} (t^a\psi)$, where $t^a$
are the standard basis of $SU(N)$ (introduced earlier) normalised as ${\rm tr}\, t^at^b
=\delta^{ab}$. $S_f^{\rm kin}$ can then be written as:
\begin{eqnarray}
S_f^{\rm kin} &=& \frac{1}{g^2}\sum_{a,b =0}^{N^2-1}\sum_{m,n =0}^{\Lambda-1}\sum_{\alpha = 1}^8\psi^{\alpha+8}_{m\,,a}K^{ab}_{mn}\psi^{\alpha}_{n\,,b}\ , \\
K_{mn}^{ab}&=&(\delta_{m+1,n} -\delta_{m,n})\delta^{ab}\pm\delta_{m,\Lambda-1}\delta_{n,0}\,d^{ab}\\
d^{ab} &=&{\rm tr}\left(t^a\,D\,t^b\,D^{\dagger}\right)\ .
\end{eqnarray}
where the plus/minus sign corresponds to periodic/ant-periodic boundary conditions. The kinetic term can also be written as:
\begin{eqnarray}
S_f^{\rm kin} &=&\sum_{a,b =0}^{N^2-1}\sum_{m,n =0}^{\Lambda-1}\sum_{\alpha,\beta = 1}^{16}\psi^{\alpha}_{m\,,a}\,{\cal M}^{\rm kin}_{mn,\alpha\beta,ab}\,\psi^{\beta}_{n\,,b}\\
{\cal M}^{\rm kin}_{mn,\alpha\beta,ab} &=&\frac{1}{2g^2}\left(\begin{array}{cc} 0_8 &  -K_{nm}^{ba}\\K_{mn}^{ab} & 0_8 \end{array}\right)_{\alpha\beta}\ .
\end{eqnarray} 
Discretising the potential part of the action is straightforward. One obtains:
\begin{eqnarray}
S_f^{\rm pot} &=&\sum_{a,b =0}^{N^2-1}\sum_{m,n =0}^{\Lambda-1}\sum_{\alpha,\beta = 1}^{16}\psi^{\alpha}_{m\,,a}\,{\cal M}^{\rm pot}_{mn,\alpha\beta,ab}\,\psi^{\beta}_{n\,,b}\\
{\cal M}^{\rm pot}_{mn,\alpha\beta,ab} &=&\frac{1}{2g^2}\,a\,if^{abc}\,(C_{9}\gamma^i)_{\alpha\beta}\,X^{c,i}_n\delta_{n+m,0}\ .
\end{eqnarray}
Finally, defining:
\begin{eqnarray}\label{M-psi}
{\cal M}_{mn,\alpha\beta,ab} &=&\frac{1}{2g^2}\left(\begin{array}{cc} 0_8 &  -K_{nm}^{ba}\\K_{mn}^{ab} & 0_8 \end{array}\right)_{\alpha\beta}+\frac{1}{2g^2}\,a\,if^{abc}\,(C_{9}\gamma^i)_{\alpha\beta}\,X^{c,i}_n\delta_{n+m,0}\ .
\end{eqnarray}
We can write:
\begin{equation}\label{S-psi}
S_f = \psi^T{\cal M}\psi\ .
\end{equation}

\subsection{The Pfaffian phase is not a problem!}
Integrating out the Fermions, the partition function of the model can be written as:
  \begin{equation}
{\cal Z} \propto\int{\cal D}X\,{\cal D}\psi\,e^{-S[X,\psi]}\propto \int{\cal D}X\,{\rm Pf}({\cal M})\,e^{-S_{\rm bos}[X]}
  \end{equation}
Observe that ${\cal M}$ is the sum of an anti-hermitian kinetic term
and a hermitian potential and ${\cal M}^\dag(X)=-{\cal M}(-X)$. 
Also because the Bosonic action is
symmetric in $X$ the Pfaffian in the partition function can be
replaced by $\vert {\rm Pf}({\cal M})\vert\cos(\Theta_{Pf})$. 
Now as long as $-\frac{\pi}{2}<\Theta_{Pf}<\frac{\pi}{2}$ the cosine is positive and the effective action defines a true probability distribution
given by 
\begin{equation}
S_{eff}=S_{\rm bos}[X]-\ln\vert {\rm Pf}({\cal M})\vert-\ln\cos(\Theta_{Pf})
\end{equation}

\begin{figure}[t] %  figure placement: here, top, bottom, or page
   \centering
   \includegraphics[width=3.9in]{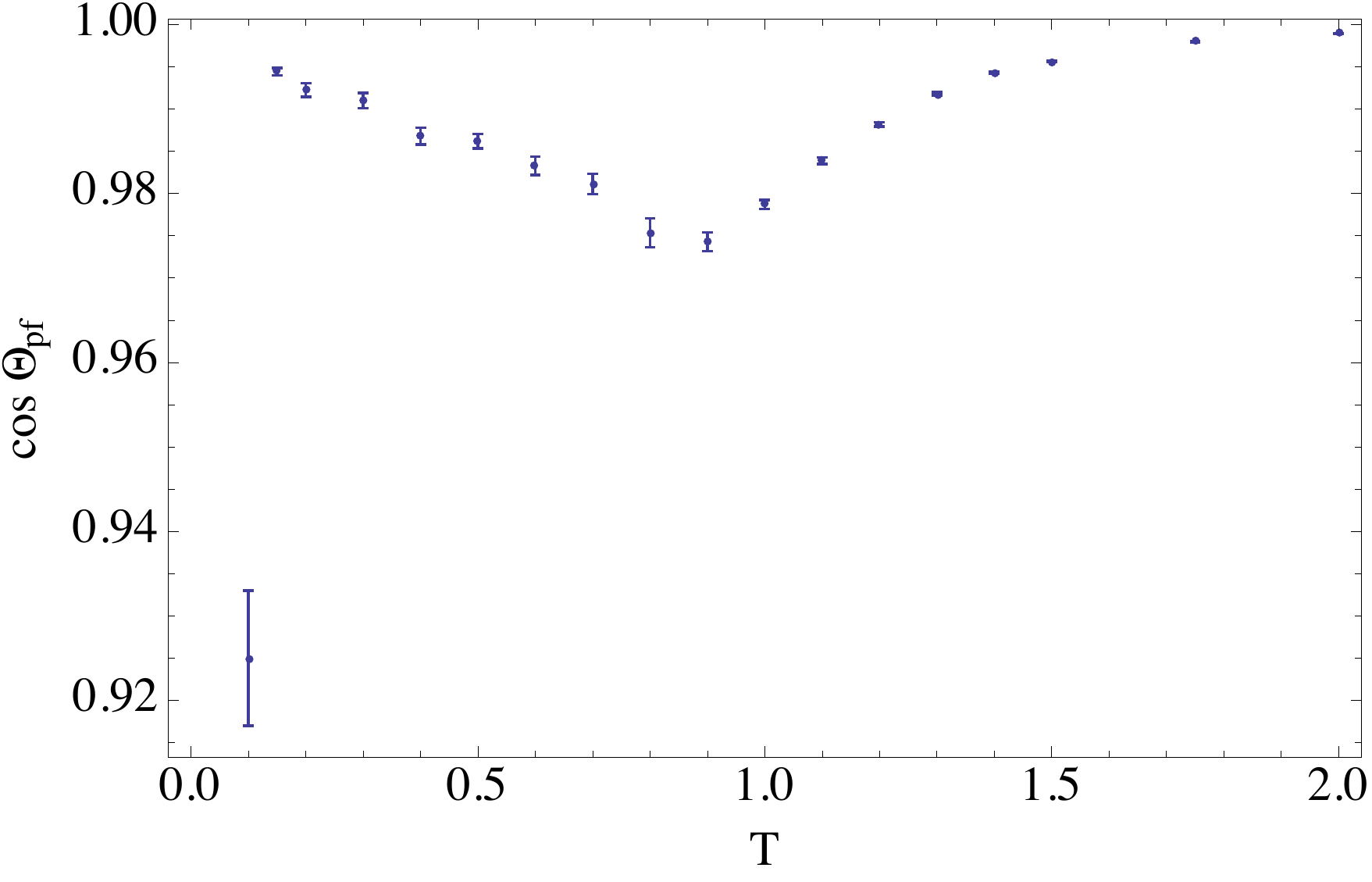} 
 \caption{\small  A plot of the phase of the pfaffian of the fermionic matrix for $N=3$ and $\Lambda = 4$. One can see that the phase remains small for all temperature, but drops at ver low temperatures. We believe that this is a lattice artifact and is not present in the continuum limit. The flat directions are controlled with a small mass regulator at low temperature.}
   \label{fig:5A}
\end{figure}

In figure \ref{fig:5A} we presented plot of the phase of the pfaffian of the fermionic matrix for $N=3$ and four lattice points.\footnote{Note that to control the flat directions at low temperature we have added a small mass for the bosonic field.} One can see that the cosine remains positive. We believe that the drop in the $\cos\theta$ curve at very low temperatures is a lattice effect and is not present in the continuum limit. Our results show a very good agreement with the earlier studies of ref.~\cite{Catterall:2009xn}.

\subsection{RHMC and fermionic forces}
  The next step is to apply the RHMC method \cite{Clark:2004cp} to the model. To this end we need the so called fermionic forces. Let us summarise briefly the philosophy.

As we have shown above the model does not suffer from a severe
sign problem and so we ignore the phase of the Pfaffian and use that:
  \begin{equation}
|{\rm Pf}({\cal M})|={\rm det}({\cal M}^{\dagger}\,{\cal M})^{1/4}\ ,
  \end{equation}
  to write 
  \begin{equation}
{\cal Z} \propto \int{\cal D}X\,{\cal D}\xi^{\dagger}\,{\cal D}\xi\,e^{-S_{\rm bos}[X]-S_{\rm ps.f}}\ ,
  \end{equation}
  where
  \begin{equation}
S_{\rm ps.f} \equiv\xi^{\dagger}\,({\cal M}^{\dagger}\,{\cal M})^{-1/4}\xi\ .
  \end{equation}
  Here $\xi$ is a $16(N_c^2-1)\Lambda$ dimensional vector consisting of the pseudo-fermionic fields. The idea of the RHMC is to approximate the rational exponent of the matrix ${\cal M}^{\dagger}\,{\cal M}$ with a partial sum:
  \begin{equation}\label{partial}
({\cal M}^{\dagger}\,{\cal M})^{\delta} =\alpha_0+\sum_{i=1}^{\#}\alpha_i\,({\cal M}^{\dagger}\,{\cal M}+\beta_i)^{-1}\ ,
  \end{equation}
  where the parameters $\alpha_0,\alpha_i,\beta_i$ and $\#$  depend on the rational exponent $\delta$, the spectral range of the matrix ${\cal M}^{\dagger}\,{\cal M}$ and the desired accuracy. We will need two rational exponents. To update the pseudo fermions we use that the field $\eta \equiv ({\cal M}^{\dagger}\,{\cal M})^{-1/8}\xi$ has a gaussian distribution and solve for $\xi = ({\cal M}^{\dagger}\,{\cal M})^{1/8}\,\eta$ using a multi-shift solver. Therefore, $\delta =1/8$ is one of the rational exponents that we need. To calculate the fermionic forces and the contribution to the hamiltonian we need to invert $({\cal M}^{\dagger}\,{\cal M})^{-1/4}$ and the second exponent is $\delta =-1/4$. 
  
  Let us elaborate on the computation of the fermionic forces. We have two type of forces: gradients with respect to $X_{n\,ij}$ and gradients with respect to the phases of the links $\theta_i$. Using the partial expansion (\ref{partial}), one can easily derive expression for the derivatives of the fermionic action:
  \begin{eqnarray}
\frac{\partial S_{\rm ps.f}}{\partial u} &=&-\sum_{i=1}^{\#}\alpha_i\,\xi^{\dagger}({\cal M}^{\dagger}\,{\cal M}+\beta_i)^{-1}\,\frac{\partial({\cal M}^{\dagger}\,{\cal M})}{\partial u}\,({\cal M}^{\dagger}\,{\cal M}+\beta_i)^{-1}\xi \nonumber \\
&&=-\sum_{i=1}^{\#}\alpha_i\,h_i^{\dagger}\,\frac{\partial({\cal M}^{\dagger}\,{\cal M})}{\partial u}\,h_i\ ,
  \end{eqnarray}
  where $h_i$ satisfy $({\cal M}^{\dagger}\,{\cal M}+\beta_i)\,h_i=\xi_i$ and are obtained from the multi-solver. 
  
  \subsection{Simulation results}
\begin{figure}[t] %  figure placement: here, top, bottom, or page
   \centering
   \includegraphics[width=3in]{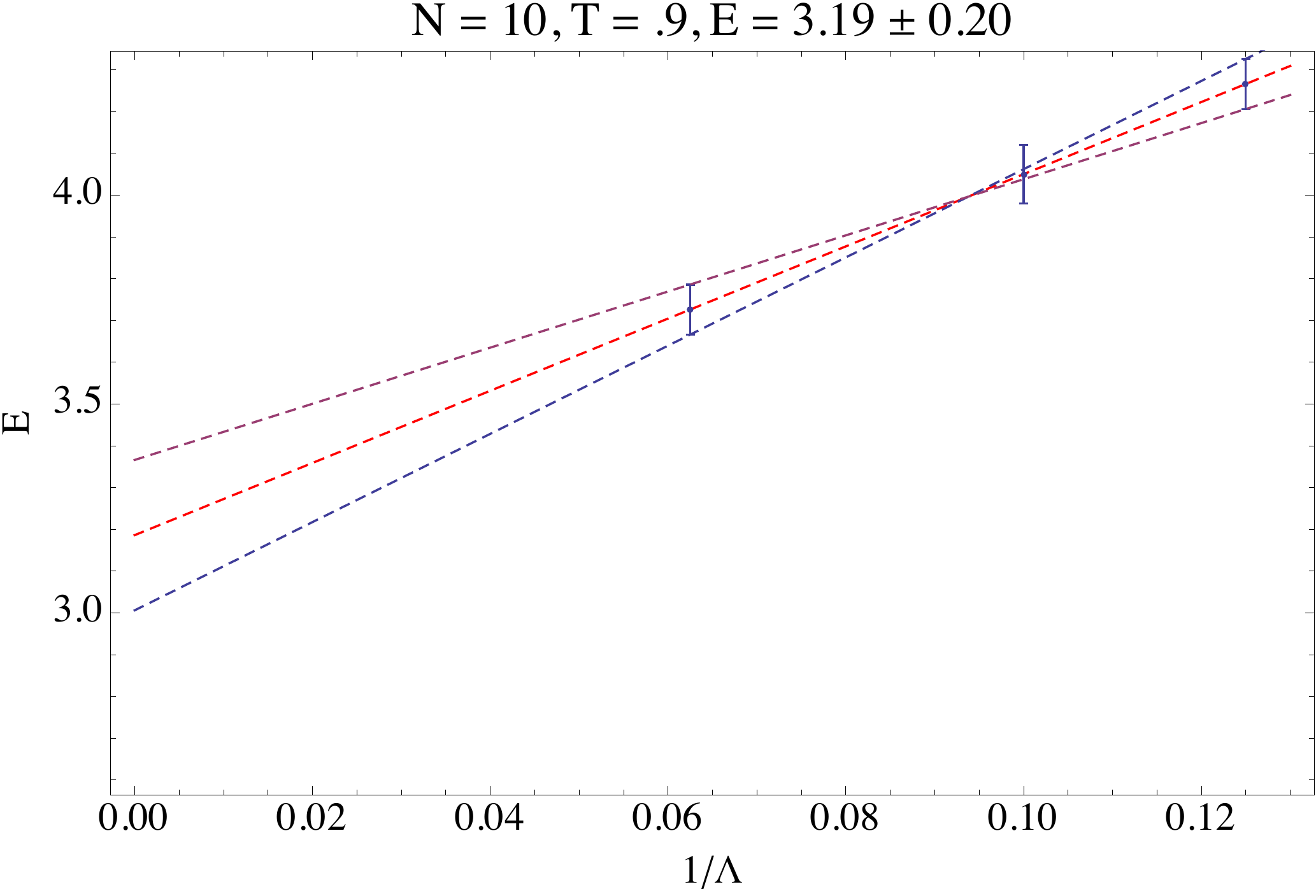} 
   \includegraphics[width=3in]{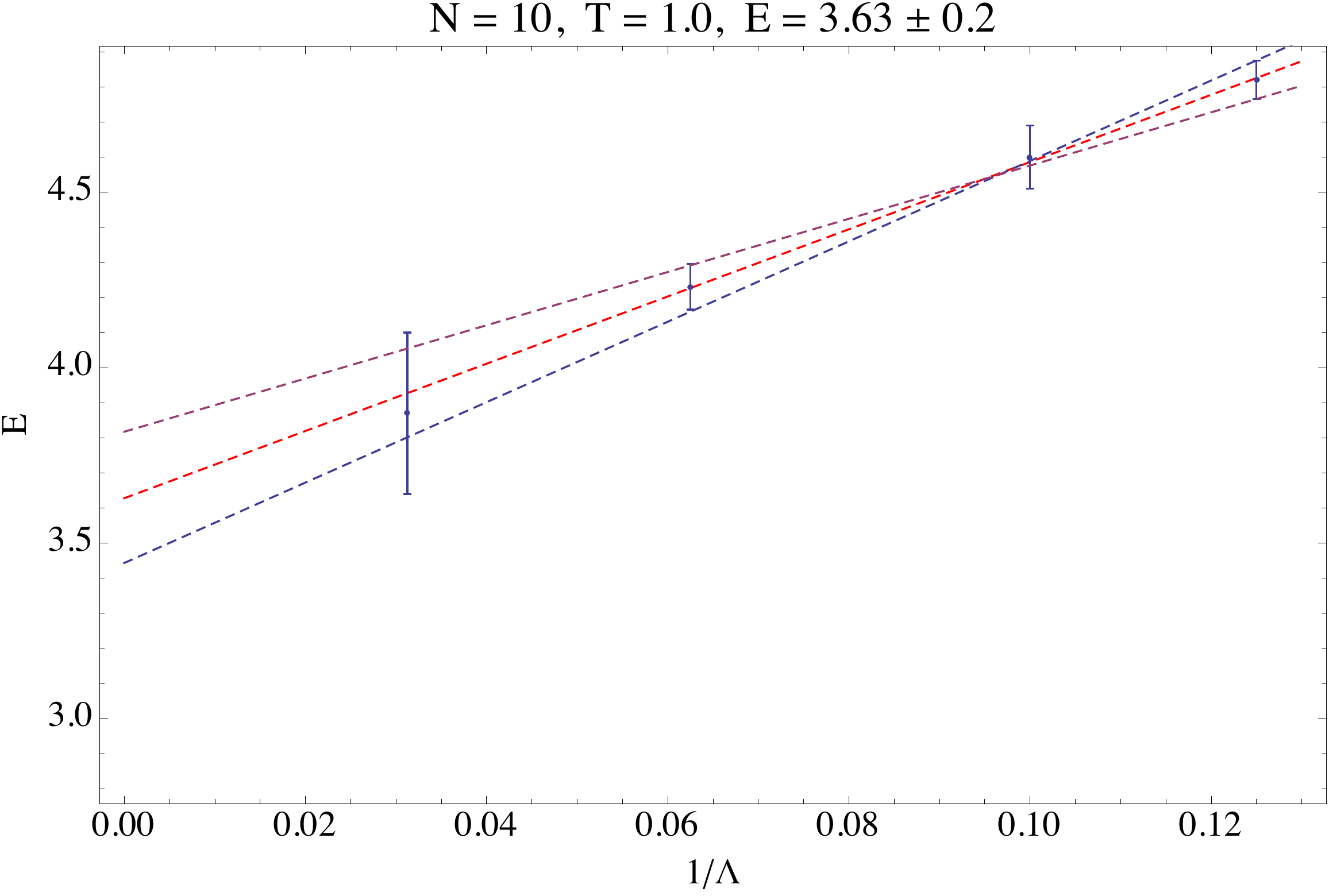} 
 \caption{\small Scaling of the internal energy with the lattice
   spacing for temperatures $T=0.9,1.0$ and $N=10$. One can see that
   the lattice effects dies our linearly, which allows extrapolation
   of the zero lattice spacing result. }
   \label{fig:6}
\end{figure}

\begin{figure}[t] %  figure placement: here, top, bottom, or page
   \centering
   \includegraphics[width=5in]{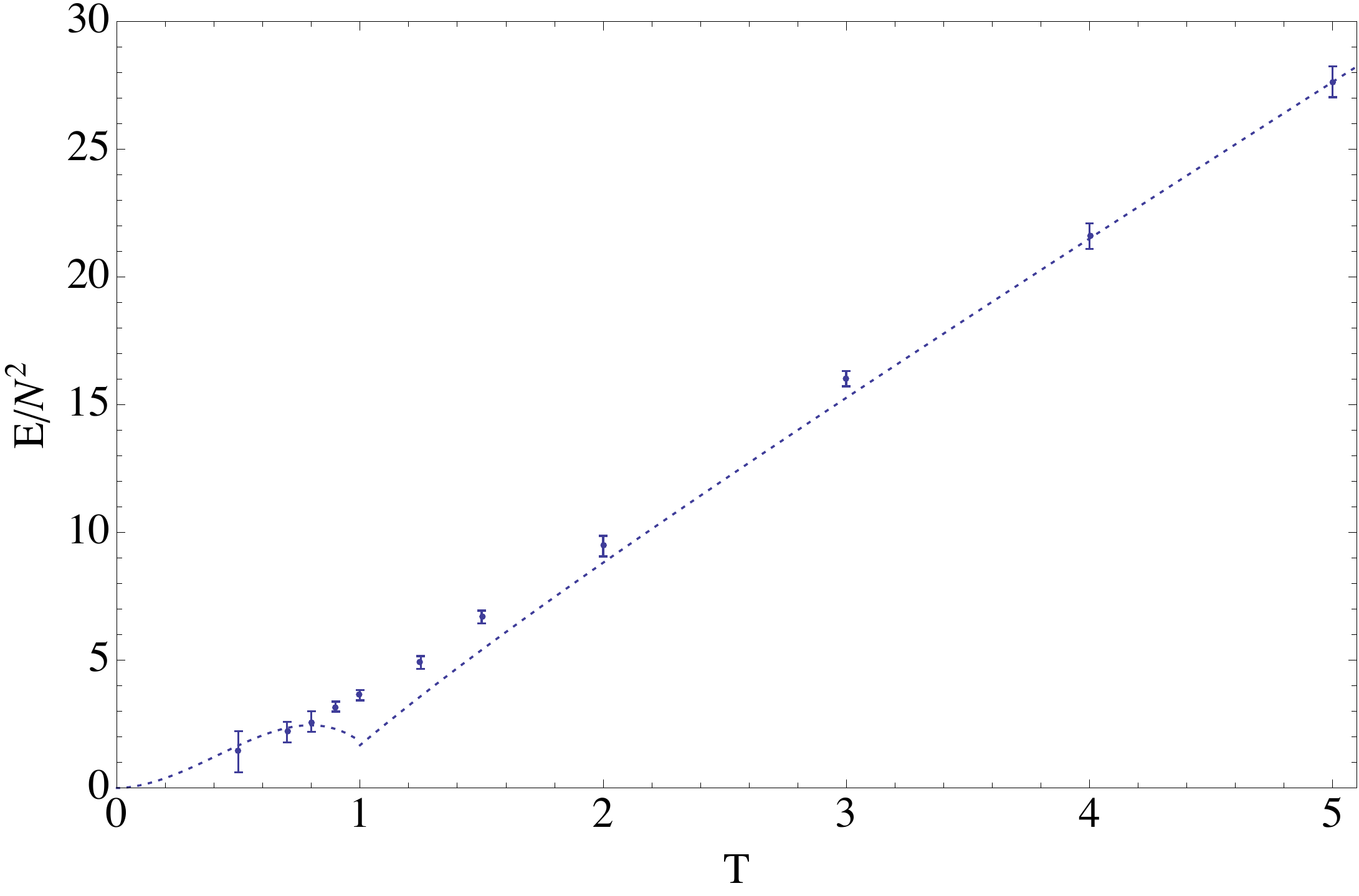} 
 \caption{\small Results for the internal energy for $8\leq N \leq 14$
   and $8\leq \Lambda \leq 16$. The dashed curve at high temperature
   correspond to the theoretical results of~\cite{Kawahara:2007ib},
   while the low temperature curve represent the prediction for the
   internal energy from the AdS/CFT correspondence.}
   \label{fig:7}
\end{figure}

In this subsection we provide our results from the Monte Carlo
simulation of the model. We focus on the same observables that we
analysed for the bosonic model in section \ref{phase structure}. The
definitions of the extent of space $\langle R^2\rangle$ and the
expectation value of the Polyakov loop $P$ remain the same. The
expression for the internal energy is
\cite{Catterall:2008yz}\footnote{Note that this expression is also
  valid for the bosonic model. The result can be obtained by rescaling
  the fields such that the kinetic term is temperature independent and
  removes any temperature dependence from the measure (the Van Vleck
  Morette determinant generically depends on temperature).  Then
  differentiating with respect to temperature and using the Ward
  identities associated with the total number of degrees of freedom
  yields this result.}:
  \begin{equation}
 \frac{ E}{N^2}=-\frac{3T}{N^2}\left(\langle S_{\rm bos}\rangle -\frac{9}{2}\Lambda\,(N^2-1)\right)\ ,
  \end{equation}
We have simulated the following configurations. For temperatures $T>2$
we have used $N=8$ and $\Lambda =8$. For the region $1\leq T\leq 2$ we
have used $N=8$ and two or three different sizes of the lattice (for
each point) in the range $8\leq\Lambda \leq 16$ (For $T=1$ we also
went to $\Lambda = 32$). For temperatures lower than one we have used
$N=10, 12, 14$ and two lattice sizes per point $\Lambda = 8, 16$.  For
all temperatures the Polyakov loop is largely independent on the
lattice spacing. The extent of space also experience very weak lattice
effects. However, this is not the case for the internal energy and
even for temperatures as high as $T = 2$ lattice effects can be a
factor. In figure \ref{fig:6} w present our results for the energy at
$T=.9,1.0$ for different lattice spacing. One can see that the lattice
effects die out linearly, which allows us to extrapolate the energy to
zero lattice spacing. Our results for the internal energy are
summarised in figure \ref{fig:7}. The dashed curve at high
temperatures is the theoretical curve obtained in the high temperature
expansion. The dashed curve at low temperature represents the
prediction of AdS/CFT.\footnote{We have used the $\alpha'$ corrected
  expression $\frac{1}{N^2}\frac{E}{\lambda^{1/3}}=
  7.41\left(\frac{T}{\lambda^{1/3}}\right)^\frac{14}{5}-5.58\left(\frac{T}{\lambda^{1/3}}\right)^\frac{23}{5}$
  obtained in~\cite{Hanada:2008ez}.} The model becomes unstable for
small matrix sizes $N$, an effect which has been related to Hawking
radiation in the dual gravitational
theory \cite{Catterall:2009xn,Hanada:2013rga}.  To compare with the
AdS/CFT predictions one needs to consider large matrices. Simulations
with large matrix sizes is computationally expensive and as a result
our low temperature data is still preliminary. However, even at this
point we have excellent overall agreement with the studies
of~\cite{Anagnostopoulos:2007fw} and \cite{Kadoh:2015mka}.

\begin{figure}[t] %  figure placement: here, top, bottom, or page
   \centering
   \includegraphics[width=3in]{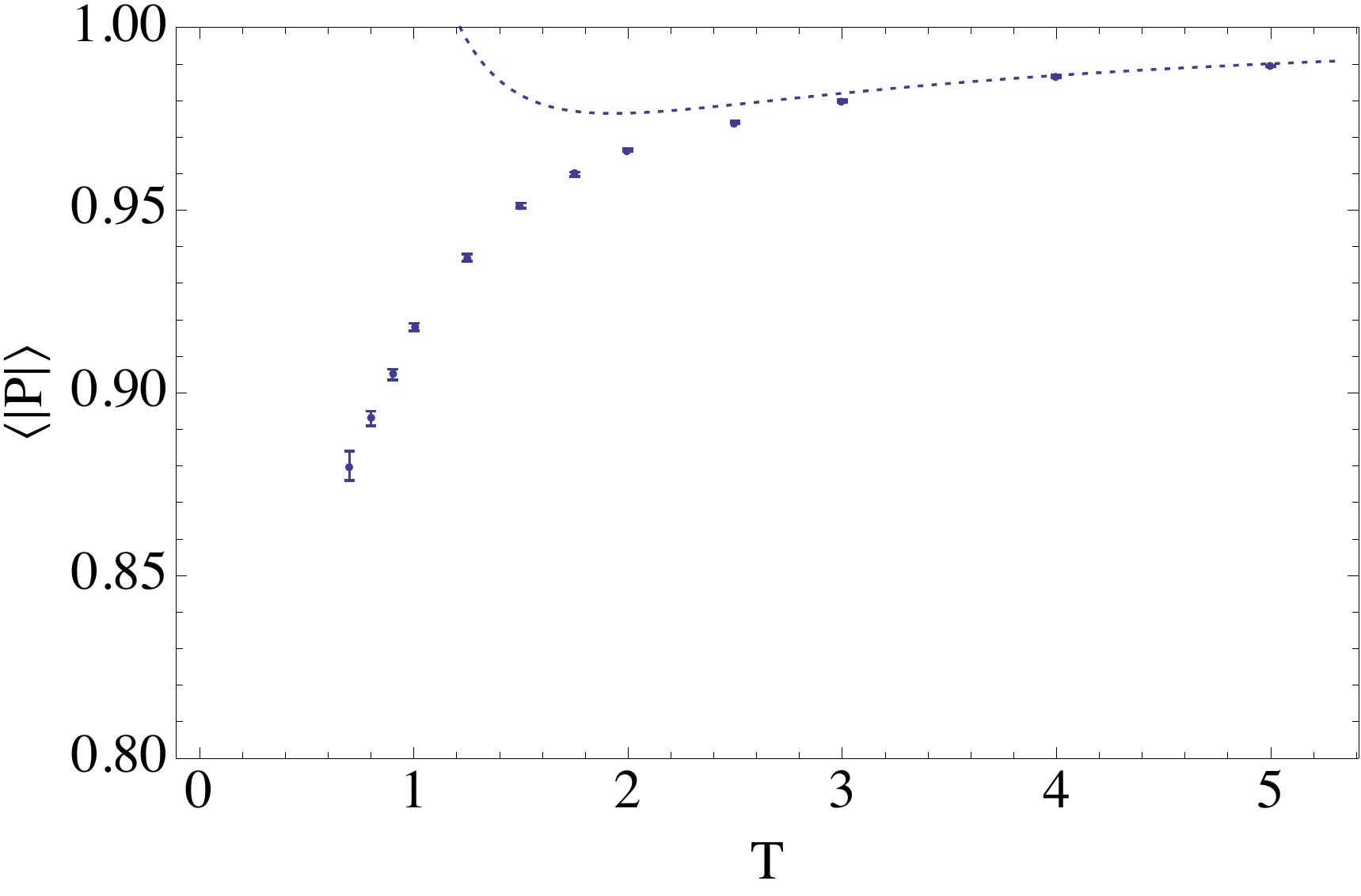} 
   \includegraphics[width=3in]{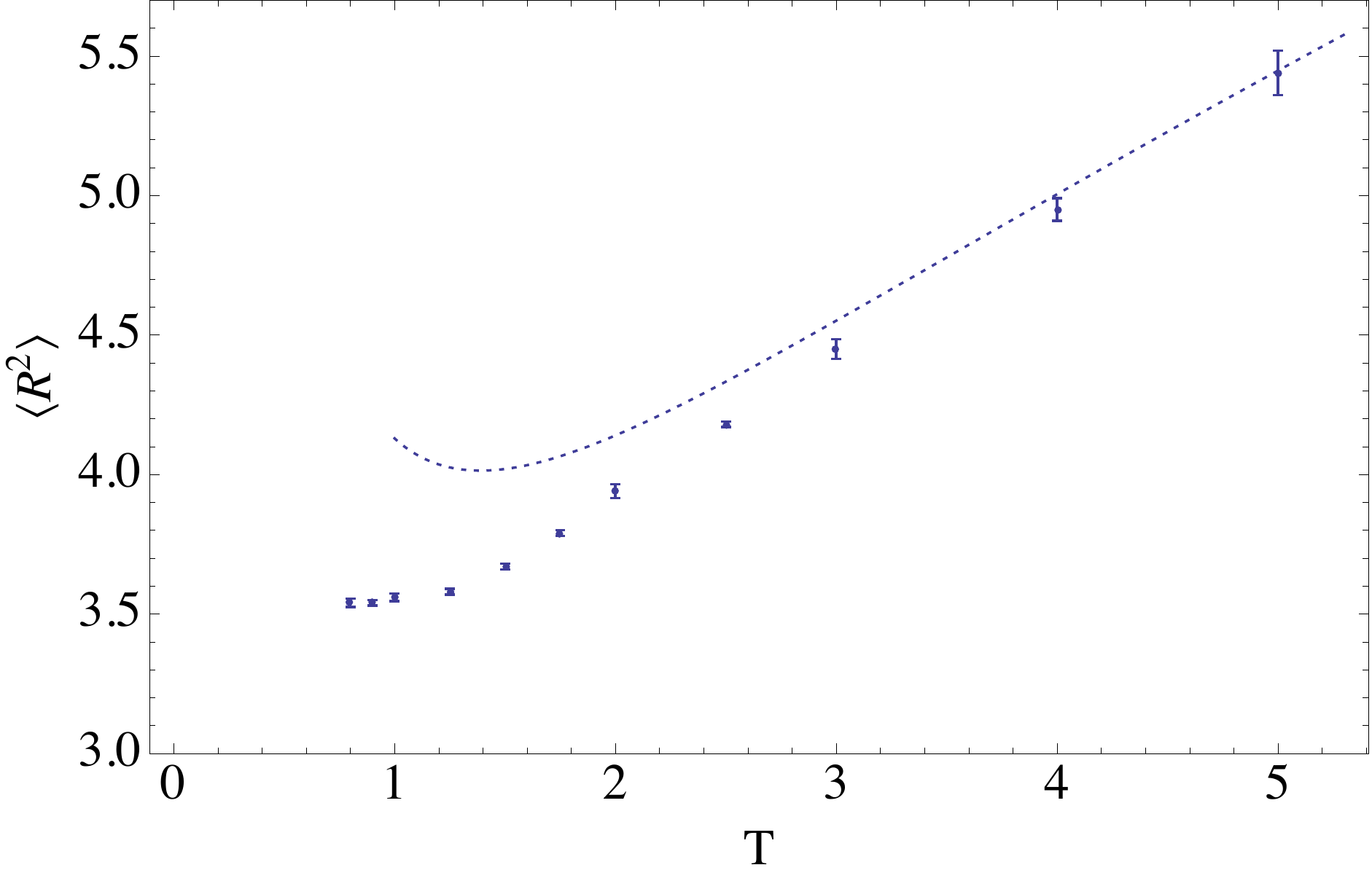} 
 \caption{\small Plots of the expectation value of the Polyakov loop
   $\langle |P|\rangle$ and the extent of space $\langle R2\rangle$ as
   functions of temperature. The dashed curves represent the
   predictions of the high temperature expansion.}
   \label{fig:8}
\end{figure}

Finally, in figure \ref{fig:8} we present our results for the Polyakov
loop and the extent of space. Again the dashed curve represents the
high temperature theoretical result obtained
in~\cite{Kawahara:2007ib}. One can see the excellent agreement at high
temperatures. Our result for these observables agree with the previous
studies preformed in~\cite{Anagnostopoulos:2007fw},
\cite{Catterall:2008yz}, \cite{Hanada:2008ez} and
\cite{Kadoh:2015mka}.
\section{Discussion}
In this paper we have analysed both the purely Bosonic and the
supersymmetric BFSS models. These are ``Hoppe'' regulated membranes
and one would expect that in the large $N$ limit when the regulator is
removed that they describe the full quantum dynamics of these
membranes.  Surprisingly we found that the bosonic model, for
sufficiently large embedding dimension reduces to a system of
$p$-massive free bosons with the mass given by $m\sim p^{1/3}$. For
$p=9$ we performed detailed simulations of the model evaluating both
the fall off of the correlation function and the eigenvalue
distribution of the $X^i$ fit with a Wigner semi-circle both of which
give a consistent mass $m\simeq (1.965\pm.007)\lambda^{1/3}$.  This is a
fundamental non-perturbative result and gives the mass gap in the full
Hamiltonian of the model. 

The correspondence of the full and gauged Gaussian model is excellent
for a wide range of temperatures.  Somewhat surprisingly the phase
transition region of the full model is faithfully reproduced by the
effective model with the two transitions of the full model merged into one.
Since the finite temperature
version of the model is also the high temperature limit of $1+1$
dimensional maximally supersymmetric Yang-Mills \cite{Aharony:2004ig}
%hep-th/0508077 and hep-th/0406210,
compactified on a circle, we have established that this latter model
should also reduce to a system of free massive scalars in its large radius
high temperature phase.

We then study the full supersymmetric BFSS model using a rational
hybrid monte carlo simulation with Fourier acceleration to evaluate
observables of the model.  After describing our lattice discretisation
of the model we investigated the phase of the Pfaffian obtained on
integrating out the Fermions. The Pfaffian is generically complex,
however, its phase is in fact not a problem for simulations. What
enters simulations is the cosine of this phase and in the
regularisation used in our work this phase is in fact restricted to a
region where the cosine is positive once the lattice spacing is
sufficiently small. Direct measurements confirm that the phase is
indeed small.

Though our results for this part of the paper do not yet go beyond
those of \cite{Anagnostopoulos:2007fw} or cover as low a temperature
as those of \cite{Kadoh:2015mka} they are more precise than those of
Catterall and Wiseman \cite{Catterall:2008yz} who use a similar
lattice simulation. We have taken several lattice spacings and then
performed an extrapolation to the limit of zero lattice spacing.  We
find good agreement with earlier results and excellent agreement with
the predictions of AdS/CFT once $1/\alpha'$ corrections are included.
Our results appear to approach the predictions of AdS/CFT a little
more closely than those of \cite{Anagnostopoulos:2007fw} but the
difference is broadly within the errors.  The principal difficulty in
simulating the model at lower temperatures is due to critical slowing
down and though Fourier acceleration helped for $T\simeq 0.5$,
simulations becomes increasingly more difficult at lower
temperatures. An additional difficulty is the instability due to flat
directions, which require increasingly larger matrix sizes as the
temperature is reduced.

One of the principal aims of this work was to check the claims of
previous work and in particular those on the absence of a complex
phase problem.  We were also interested in calibrating our code as we
extend it to include systems with $D4$-branes. The extension to such
systems will allow us to perform more extensive tests of the AdS/CFT
correspondence.

{\bf Acknowledgements:} We wish to thank S. Catterall, M. Panero and
M. Hanada for helpful comments.


\begin{thebibliography}{99}

%\cite{Townsend:1995kk}
\bibitem{Townsend:1995kk}
  P.~K.~Townsend,
  ``The eleven-dimensional supermembrane revisited'',
  {\it Phys. Lett.} {\bf B350} (1995) 184
  [hep-th/9501068].
  %%CITATION = HEP-TH/9501068;%%
  
  %\cite{Banks:1996vh}
\bibitem{Banks:1996vh} 
  T.~Banks, W.~Fischler, S.~H.~Shenker and L.~Susskind,
  ``M theory as a matrix model: A Conjecture,''
  Phys.\ Rev.\ D {\bf 55}, 5112 (1997)
  [hep-th/9610043].
  %%CITATION = HEP-TH/9610043;%%
  %2360 citations counted in INSPIRE as of 19 May 2015
  
  %\cite{Susskind:1997cw}
\bibitem{Susskind:1997cw}
  L.~Susskind,
  %``Another conjecture about M(atrix) theory,''
  hep-th/9704080.
  %%CITATION = HEP-TH/9704080;%%
  %395 citations counted in INSPIRE as of 03 juin 2015
  
  %\cite{Berenstein:2003gb}
\bibitem{Berenstein:2003gb}
  D.~E.~Berenstein, J.~M.~Maldacena and H.~S.~Nastase,
  ``Strings in flat space and pp waves from N=4 Super Yang Mills,''
  AIP Conf.\ Proc.\  {\bf 646} (2003) 3.
  %%CITATION = APCPC,646,3;%%

%\cite{Aharony:2008gk}
\bibitem{Aharony:2008gk}
  O.~Aharony, O.~Bergman and D.~L.~Jafferis,
  ``Fractional M2-branes,''
  JHEP {\bf 0811} (2008) 043
  [arXiv:0807.4924 [hep-th]].
  %%CITATION = ARXIV:0807.4924;%%

%\cite{Hoppe:PhDThesis1982}
\bibitem{Hoppe:PhDThesis1982} 
  J.~Hoppe 
  ``Quantum Theory Of A Massless Relativistic Surface And A Two Dimensional
  Bound State Problem'', 
  Ph.D. Thesis, Massachusetts Institute of Technology, (1982).

%\cite{de Wit:1988ig}
\bibitem{de Wit:1988ig}
  B.~de Wit, J.~Hoppe and H.~Nicolai,
  ``On the Quantum Mechanics of Supermembranes,''
  Nucl.\ Phys.\ B {\bf 305} (1988) 545.
  %%CITATION = NUPHA,B305,545;%%
  
%\cite{Mandal:2011hb} 
\bibitem{Mandal:2011hb}
  G.~Mandal and T.~Morita,
  ``Phases of a two dimensional large N gauge theory on a torus,''
  Phys.\ Rev.\ D {\bf 84} (2011) 085007
  doi:10.1103/PhysRevD.84.085007
  [arXiv:1103.1558 [hep-th]].
  %%CITATION = doi:10.1103/PhysRevD.84.085007;%%

%\cite{Mandal:2009vz}
\bibitem{Mandal:2009vz} 
  G.~Mandal, M.~Mahato and T.~Morita,
  ``Phases of one dimensional large N gauge theory in a 1/D expansion,''
  JHEP {\bf 1002}, 034 (2010)
  doi:10.1007/JHEP02(2010)034
  [arXiv:0910.4526 [hep-th]].
  %%CITATION = doi:10.1007/JHEP02(2010)034;%%
  %21 citations counted in INSPIRE as of 15 Mar 2016

%\cite{Azuma:2014cfa}
\bibitem{Azuma:2014cfa}
  T.~Azuma, T.~Morita and S.~Takeuchi,
  ``Hagedorn Instability in Dimensionally Reduced Large-N Gauge Theories as Gregory-Laflamme and Rayleigh-Plateau Instabilities,''
  Phys.\ Rev.\ Lett.\  {\bf 113} (2014) 091603
  doi:10.1103/PhysRevLett.113.091603
  [arXiv:1403.7764 [hep-th]].
  %%CITATION = doi:10.1103/PhysRevLett.113.091603;%%
  %4 citations counted in INSPIRE as of 11 Mar 2016
  
%\cite{Catterall:2010fx}
\bibitem{Catterall:2010fx}
   S.~Catterall, A.~Joseph and T.~Wiseman,
   ``Thermal phases of D1-branes on a circle from lattice super Yang-Mills,''
   JHEP {\bf 1012}, 022 (2010)
   [arXiv:1008.4964 [hep-th]].

%\cite{Hiller:2005vf}
\bibitem{Hiller:2005vf} 
  J.~R.~Hiller, S.~S.~Pinsky, N.~Salwen and U.~Trittmann,
  %``Direct evidence for the Maldacena conjecture for N=(8,8) super Yang-Mills theory in 1+1 dimensions,''
  Phys.\ Lett.\ B {\bf 624}, 105 (2005)
  doi:10.1016/j.physletb.2005.08.003
  [hep-th/0506225].
  %%CITATION = doi:10.1016/j.physletb.2005.08.003;%%
  %15 citations counted in INSPIRE as of 01 May 2016

%\cite{Baake:1984ie}
\bibitem{Baake:1984ie}
  M.~Baake, M.~Reinicke and V.~Rittenberg,
  ``Fierz Identities for Real Clifford Algebras and the Number of Supercharges,''
  J.\ Math.\ Phys.\  {\bf 26} (1985) 1070.
  %%CITATION = JMAPA,26,1070;%%
  %100 citations counted in INSPIRE as of 02 juin 2015

%\cite{Flume:1984mn}
\bibitem{Flume:1984mn}
  R.~Flume,
  ``On Quantum Mechanics With Extended Supersymmetry and Nonabelian Gauge Constraints,''
  Annals Phys.\  {\bf 164} (1985) 189.
  %%CITATION = APNYA,164,189;%%
  %97 citations counted in INSPIRE as of 02 juin 2015

%\cite{Claudson:1984th}
\bibitem{Claudson:1984th}
  M.~Claudson and M.~B.~Halpern,
  ``Supersymmetric Ground State Wave Functions,''
  Nucl.\ Phys.\ B {\bf 250} (1985) 689.
  %%CITATION = NUPHA,B250,689;%%
  %246 citations counted in INSPIRE as of 02 juin 2015


%\cite{Witten:1995im}
\bibitem{Witten:1995im}
  E.~Witten,
  ``Bound states of strings and p-branes,''
  Nucl.\ Phys.\ B {\bf 460} (1996) 335
  [hep-th/9510135].
  %%CITATION = HEP-TH/9510135;%%
  %1288 citations counted in INSPIRE as of 02 juin 2015
  
%\cite{Kaplan:2005ta}
\bibitem{Kaplan:2005ta}
  D.~B.~Kaplan and M.~Unsal,
  ``A Euclidean lattice construction of supersymmetric Yang-Mills theories with sixteen supercharges,''
  JHEP {\bf 0509} (2005) 042
  doi:10.1088/1126-6708/2005/09/042
  [hep-lat/0503039].
  %%CITATION = doi:10.1088/1126-6708/2005/09/042;%%
  %122 citations counted in INSPIRE as of 11 Mar 2016

  %\cite{Catterall:2008yz}
\bibitem{Catterall:2008yz} 
  S.~Catterall and T.~Wiseman,
  ``Black hole thermodynamics from simulations of lattice Yang-Mills theory,''
  Phys.\ Rev.\ D {\bf 78}, 041502 (2008)
  [arXiv:0803.4273 [hep-th]].
  %%CITATION = ARXIV:0803.4273;%%
  %87 citations counted in INSPIRE as of 28 May 2015

%\cite{Anagnostopoulos:2007fw}
\bibitem{Anagnostopoulos:2007fw} 
  K.~N.~Anagnostopoulos, M.~Hanada, J.~Nishimura and S.~Takeuchi,
  ``Monte Carlo studies of supersymmetric matrix quantum mechanics with sixteen supercharges at finite temperature,''
  Phys.\ Rev.\ Lett.\  {\bf 100}, 021601 (2008)
  [arXiv:0707.4454 [hep-th]].
  %%CITATION = ARXIV:0707.4454;%%
  %111 citations counted in INSPIRE as of 28 May 2015
  
  %\cite{Hanada:2008ez}
\bibitem{Hanada:2008ez} 
  M.~Hanada, Y.~Hyakutake, J.~Nishimura and S.~Takeuchi,
  %``Higher derivative corrections to black hole thermodynamics from supersymmetric matrix quantum mechanics,''
  Phys.\ Rev.\ Lett.\  {\bf 102}, 191602 (2009)
  [arXiv:0811.3102 [hep-th]].
  %%CITATION = ARXIV:0811.3102;%%
  %69 citations counted in INSPIRE as of 03 juin 2015
  
  %\cite{Kadoh:2015mka}
\bibitem{Kadoh:2015mka} 
  D.~Kadoh and S.~Kamata,
  ``Gauge/gravity duality and lattice simulations of one dimensional SYM with sixteen supercharges,''
  arXiv:1503.08499 [hep-lat].
  %%CITATION = ARXIV:1503.08499;%%

  %\cite{Kawahara:2007fn}
\bibitem{Kawahara:2007fn} 
  N.~Kawahara, J.~Nishimura and S.~Takeuchi,
  ``Phase structure of matrix quantum mechanics at finite temperature,''
  JHEP {\bf 0710}, 097 (2007)
  [arXiv:0706.3517 [hep-th]].
  %%CITATION = ARXIV:0706.3517;%%
  %30 citations counted in INSPIRE as of 19 May 2015

%\cite{Polchinski} 
\bibitem{Polchinski}  
Polchinski, Joseph. "Frontmatter", String Theory. 1st ed. Vol. 2. Cambridge: Cambridge University Press, 1998. pp. i-viii. Cambridge Books Online.
  
%\cite{Aharony:2004ig}
\bibitem{Aharony:2004ig}
  O.~Aharony, J.~Marsano, S.~Minwalla and T.~Wiseman,
  ``Black hole-black string phase transitions in thermal 1+1 dimensional supersymmetric Yang-Mills theory on a circle,''
  Class.\ Quant.\ Grav.\  {\bf 21} (2004) 5169
  [hep-th/0406210].
  %%CITATION = HEP-TH/0406210;%%
  %110 citations counted in INSPIRE as of 02 Jun 2015  
  
  %\cite{Aharony:2005ew}
\bibitem{Aharony:2005ew} 
  O.~Aharony, J.~Marsano, S.~Minwalla, K.~Papadodimas, M.~Van Raamsdonk and T.~Wiseman,
  %``The Phase structure of low dimensional large N gauge theories on Tori,''
  JHEP {\bf 0601}, 140 (2006)
  doi:10.1088/1126-6708/2006/01/140
  [hep-th/0508077].
  %%CITATION = doi:10.1088/1126-6708/2006/01/140;%%
  %63 citations counted in INSPIRE as of 01 May 2016
  
    %\cite{Kawahara:2007ib}
\bibitem{Kawahara:2007ib} 
  N.~Kawahara, J.~Nishimura and S.~Takeuchi,
  ``High temperature expansion in supersymmetric matrix quantum mechanics,''
  JHEP {\bf 0712}, 103 (2007)
  [arXiv:0710.2188 [hep-th]].
  %%CITATION = ARXIV:0710.2188;%%
  %22 citations counted in INSPIRE as of 19 May 2015

  %\cite{Gross:1980he}
\bibitem{Gross:1980he} 
  D.~J.~Gross and E.~Witten,
  ``Possible Third Order Phase Transition in the Large N Lattice Gauge Theory,''
  Phys.\ Rev.\ D {\bf 21}, 446 (1980).
  %%CITATION = PHRVA,D21,446;%%
  %592 citations counted in INSPIRE as of 20 May 2015
  S.~R.~Wadia,
  ``$N$ = Infinity Phase Transition in a Class of Exactly Soluble Model Lattice Gauge Theories,''
  Phys.\ Lett.\ B {\bf 93}, 403 (1980).
  %%CITATION = PHLTA,B93,403;%%
  %133 citations counted in INSPIRE as of 20 May 2015
  
  %\cite{Hotta:1998en}
\bibitem{Hotta:1998en} 
  T.~Hotta, J.~Nishimura and A.~Tsuchiya,
  ``Dynamical aspects of large N reduced models,''
  Nucl.\ Phys.\ B {\bf 545}, 543 (1999)
  [hep-th/9811220].
  %%CITATION = HEP-TH/9811220;%%
  %109 citations counted in INSPIRE as of 22 May 2015
  
  %\cite{Brezin:1977sv}
\bibitem{Brezin:1977sv} 
  E.~Brezin, C.~Itzykson, G.~Parisi and J.~B.~Zuber,
  ``Planar Diagrams,''
  Commun.\ Math.\ Phys.\  {\bf 59}, 35 (1978).
  %%CITATION = CMPHA,59,35;%%
  %1073 citations counted in INSPIRE as of 22 May 2015

%\cite{Catterall:2009xn}
\bibitem{Catterall:2009xn}
  S.~Catterall and T.~Wiseman,
  %``Extracting black hole physics from the lattice,''
  JHEP {\bf 1004} (2010) 077
  [arXiv:0909.4947 [hep-th]].
  %%CITATION = ARXIV:0909.4947;%%
  %48 citations counted in INSPIRE as of 02 juin 2015
  
   %\cite{Clark:2004cp}
\bibitem{Clark:2004cp} 
  M.~A.~Clark, A.~D.~Kennedy and Z.~Sroczynski,
  ``Exact 2+1 flavour RHMC simulations,''
  Nucl.\ Phys.\ Proc.\ Suppl.\  {\bf 140}, 835 (2005)
  [hep-lat/0409133].
  %%CITATION = HEP-LAT/0409133;%%
  %89 citations counted in INSPIRE as of 24 Apr 2015

%\cite{Hanada:2013rga}
\bibitem{Hanada:2013rga}
  M.~Hanada, Y.~Hyakutake, G.~Ishiki and J.~Nishimura,
  %``Holographic description of quantum black hole on a computer,''
  Science {\bf 344} (2014) 882
  [arXiv:1311.5607 [hep-th]].
  %%CITATION = ARXIV:1311.5607;%%
  %25 citations counted in INSPIRE as of 02 juin 2015

  \end{thebibliography}
\end{document}